\documentclass[sigconf]{acmart}

\AtBeginDocument{%
  }

\copyrightyear{2024}
\acmYear{2024}
\setcopyright{rightsretained}
\acmConference[E-Energy '24]{The 15th ACM International Conference on Future and Sustainable Energy Systems}{June 4--7, 2024}{Singapore, Singapore}
\acmBooktitle{The 15th ACM International Conference on Future and Sustainable Energy Systems (E-Energy '24), June 4--7, 2024, Singapore, Singapore}
\acmDOI{10.1145/3632775.3661947}
\acmISBN{979-8-4007-0480-2/24/06}

\usepackage[T1]{fontenc}
\usepackage{mathtools}
\usepackage{textcomp}
\mathtoolsset{centercolon}

\usepackage{hyperref}
\hypersetup{
    colorlinks=false,
    linkcolor=blue,
    filecolor=magenta,      
    urlcolor=cyan,
    pdfpagemode=FullScreen,
    }
\newcommand{\vgubar}{\overline{\vct{g}}}

\newcommand{\R}{\mathbb{R}}

\newcommand{\vct}[1]{\boldsymbol{#1}}
\newcommand{\mtx}[1]{\boldsymbol{#1}}

\newcommand{\<}{\langle}
\renewcommand{\>}{\rangle}

\newcommand{\T}{\top}

\newcommand{\ip}[2]{\left\<#1, #2\right\>}

\renewcommand{\c}[1]{\left\{#1\right\}}

\newcommand{\set}[1]{\mathcal{#1}}

\newcommand{\linop}[1]{\mathcal{#1}}    

\DeclareMathOperator*{\minimize}{\operatorname{minimize}}

\newcommand{\vb}{\vct{b}}

\newcommand{\vd}{\vct{d}}
\newcommand{\ve}{\vct{e}}

\newcommand{\vg}{\vct{g}}
\newcommand{\vh}{\vct{h}}

\newcommand{\vk}{\vct{k}}

\newcommand{\vp}{\vct{p}}

\newcommand{\vs}{\vct{s}}

\newcommand{\vu}{\vct{u}}
\newcommand{\vw}{\vct{w}}
\newcommand{\vx}{\vct{x}}
\newcommand{\vy}{\vct{y}}
\newcommand{\vz}{\vct{z}}
\newcommand{\valpha}{\vct{\alpha}}
\newcommand{\vbeta}{\vct{\beta}}

\newcommand{\vtheta}{\vct{\theta}}

\newcommand{\vlambda}{\vct{\lambda}}
\newcommand{\vmu}{\vct{\mu}}
\newcommand{\vnu}{\vct{\nu}}

\newcommand{\vpi}{\vct{\pi}}
\newcommand{\vrho}{\vct{\rho}}

\newcommand{\vomega}{\vct{\omega}}
\newcommand{\vzero}{\vct{0}}
\newcommand{\vone}{\vct{1}}

\newcommand{\mA}{\mtx{A}}
\newcommand{\mB}{\mtx{B}}

\newcommand{\mF}{\mtx{F}}
\newcommand{\mG}{\mtx{G}}

\newcommand{\mQ}{\mtx{Q}}

\newcommand{\mPhi}{\mtx{\Phi}}

\newcommand{\mId}{\mathbf{I}}

\newcommand{\loL}{\linop{L}}

\newcommand{\setD}{\set{D}}

\newcommand{\setN}{\set{N}}
\newcommand{\setO}{\set{O}}
\newcommand{\setP}{\set{P}}

\newcommand{\setT}{\set{T}}
\newcommand{\setU}{\set{U}}

\newcommand{\diag}{\operatorname{diag}}

\newcommand{\vpubar}{\overline{\vp}}
\newcommand{\DBDD}{\frac{\partial \vb}{\partial \vd}}

\AtEndPreamble{%
\theoremstyle{acmdefinition}

\newtheorem{assumption}[theorem]{Assumption}
}

\DeclareSymbolFont{sfoperators}{OT1}{cmss}{m}{n}
\DeclareSymbolFontAlphabet{\mathsf}{sfoperators}

\makeatletter
\def\operator@font{\mathgroup\symsfoperators}
\makeatother

\begin{document}

\title[Locational marginal burden]{Locational marginal burden:\\ Quantifying the equity of optimal power flow solutions}


\author{Samuel Talkington}
\authornote{Both authors contributed equally to this research.}
\affiliation{%
  \institution{Georgia Institute of Technology}
  \city{Atlanta}
  \state{Georgia}
  \country{USA}}
\email{talkington@gatech.edu}
\orcid{0000-0001-5768-8115}

\author{Amanda West}
\authornotemark[1]
\affiliation{%
    \institution{Georgia Institute of Technology}
    \city{Atlanta}
    \state{Georgia}
    \country{USA}}
\email{awest93@gatech.edu}
\orcid{0000-0002-5236-3589}

\author{Rabab Haider}
\affiliation{%
  \institution{Georgia Institute of Technology}
  \city{Atlanta}
  \state{Georgia}
  \country{USA}}
\email{rabab.haider@gatech.edu}
\orcid{0000-0002-5409-9769}

\renewcommand{\shortauthors}{Talkington, West, Haider}

\begin{abstract}
 Fair distribution of benefits in electric power systems is a pertinent energy policymaking problem; however, these efforts cannot be easily quantified in power system engineering studies.  Therefore, we propose \textit{locational marginal burden} (LMB) to provide an interface between a well-studied measure of energy pricing equity\textemdash energy burden\textemdash with an optimal power flow problem (OPF). This is achieved by investigating the intrinsic link between the dual optimal solution of an OPF problem and the electricity prices, which are used to calculate the energy burden. By applying results from the field of differentiable optimization, locational marginal prices (LMPs) associated with an OPF solution can be differentiated with respect to demand. This enables electricity retail prices\textemdash and thereby, energy burden itself\textemdash to be differentiated, resulting in the proposed LMB. Simulation of a synthetic Hawaii network interfaced with real-world socioeconomic data shows how the LMB provides new insights into how the operation of the electricity network affects the equity of energy prices.
\end{abstract}



\received{2 February 2024}
\received[accepted]{12 March 2024}

\maketitle

\newpage

\section{Introduction}\label{sec:intro}

Access to electricity is essential for life in the modern world. Unfortunately, some have easier access than others. The way in which the infrastructure of the electric power system is chosen to be designed and managed can have a significant influence on whether access to energy is distributed equitably. The ongoing challenges to ensure equity in the operation of the electric power system have been exemplified by the proliferation of distributed energy resources \cite{brockway_inequitable_2021}, distribution network reconfiguration \cite{taylor_managing_wildfire_equity_2023}, power shut-offs during extreme events \cite{Kody_sharing_2022,sundar2023fairly}, decarbonization methodologies \cite{wamburu_data_driven_equitable_decarbonization_2022,lechowicz_equitable_decarbonization_2023}, and many other broad social challenges \cite{ansarin_review_2022}. 

An overarching goal within the energy equity landscape is to reduce systemic barriers to reliable and affordable access to electricity \cite{horowitz_equity_2014}. This area of focus is receiving increased interest \cite{balogun_equitable_pricing_2023,chen_retail_equity_2023} due to ongoing changes in the design of the retail electricity market \cite{haider_reinventing_2021}, which can increasingly lead to more dynamic pricing mechanisms for end users. Simultaneously, global and national initiatives to combat climate change, improve energy affordability, and ensure equity in access to reliable electricity may lead to significant changes in the way customers are billed for electricity \cite{J40_EO14008_sec223}.

Commonly, equity in energy affordability is assessed by the \textit{energy burden} metric, which  is defined as the fraction of a household's income spent purchasing energy \cite{ACEEEEnergyAffordability}. Although the energy burden metric encompasses all forms of energy, it is often discussed in the context of access to affordable \textit{electricity}. Previous studies have characterized household energy burden at the building scale or at the census level \cite{pittman_energy_2023, scheier_measurement_2022,chen_localized_2022, memmott_sociodemographic_2021,charlier_energy_2018}. These studies are typically conducted post hoc by collecting census, customer, and pricing data to calculate energy burden. While they provide immense insight, existing energy burden studies are often focused on policymaking. The existing literature has thus far produced minimal investigations into how the \textit{underlying electricity network operations and investments affect energy burden}. Moreover, there have not been any studies directly relating energy burden and the underlying grid infrastructure, which supplies electricity to customers. 

Whereas energy burden represents an equity metric for individual customers in relation to their income and electricity affordability, we propose the locational marginal burden (LMB) as a metric to analyze energy burden within the \emph{operational context} of the power system. The proposed LMB metric evaluates the inherent link between energy affordability and an optimal power flow (OPF) solution via the electricity prices, which are related to the dual optimal solution of an OPF problem. Analyzing the problem of inequitable energy burden within power system operations may lead to novel methods and programs for remediating this burden at both the  transmission and distribution levels. 


Therefore, this work aims to close the gap between energy burden and electricity network operations by creating\textemdash for the first time\textemdash an explicit link between energy burden and the optimal power flow (OPF) problem.  This results in a computational tool that we term \emph{locational marginal burden}, which comprises the sensitivity of energy burden with respect to the change in demand at each transmission node in a network. We hypothesize that this metric may be used by both engineers and policymakers for assessing, predicting, and improving the equity of optimal power flow solutions within the areas they serve.

\subsection{Background and related work}
\label{sec:background}

\subsubsection{Equity in electric power systems}

    Equity in residential electricity pricing is a topic that has received significant research interest \cite{horowitz_equity_2014}. The authors of \cite{balogun_equitable_pricing_2023} used differentiable optimization to solve a related problem in pricing equity; they determined LMPs that minimize a weighted combination of voltage deviations and a proposed inequity metric for the entire network. The method requires manually tuning trade-off parameters and does not use socioeconomic data as is typically required in energy burden metrics \cite{ACEEEEnergyAffordability,pittman_energy_2023}.

    Quantifying the equity of electricity pricing schemes, as well as embedding the trade-off between equity and efficiency in the algorithms which determine the design of electricity rates, is not new. This topic dates back to at least \cite{naughton_efficiency_1986}, and was contemporarily reviewed in \cite{ansarin_review_2022}. Recent research \cite{chen_retail_equity_2023,balogun_equitable_pricing_2023} quantitatively incorporates equity metrics into grid-aware optimization problems via both a bi-level formulation \cite{chen_retail_equity_2023} and an iterative subgradient method \cite{balogun_equitable_pricing_2023}. Importantly, these methods develop tailor-made optimization algorithms for enforcing equity goals; however, these methods do not directly quantify how solutions to conventional optimal power flow solutions impact equity metrics. 

  The power system community has begun investigating equity concerns in system planning, operations, and resource integration. At the transmission level, research into system planning has focused on equitable repair of transmission lines, undergrounding lines to mitigate the impacts of extreme weather events, and leveraging distribution-level resources to supplement and replace transmission generation \cite{Kody2022OptimizingTI, roald_power_2023, roald_chance-constrained_2018}. A second body of literature considers power systems operations to alleviate risks to under-resourced communities during various grid events, with a particular focus on managing and mitigating negative impacts of wildfires \cite{Kody_sharing_2022,taylor_managing_wildfire_equity_2023}. At the distribution level, the authors of \cite{brockway_inequitable_2021} investigate the inequities in the access and adoption of distributed energy resources, and adoption limits imposed by possible systemic inequities in the distribution power infrastructure. However, to our knowledge, prior literature has not analyzed the impact of transmission power flows on the nodal energy burden. Our work fills this gap by assessing the impact of system and market operators' actions on energy burden. This analysis would provide information for accurately utilizing new funding opportunities. For example, in the United States, recent funding mandates seek to ensure 40\% of all energy infrastructure funding goes to under-resourced communities \cite{J40_EO14008_sec223}; moreover, transmission capacity expansion is one of the Biden administration's priorities and is implemented through the Inflation Reduction Act—Transmission Siting and Economic Development Program.

\subsubsection{Marginal pricing and emissions}
The locational marginal prices (LMPs) in a transmission network are defined as the change in total operating cost incurred for serving one additional unit of demand. In other words, the LMPs can be obtained by taking the gradient of the Lagrangian function of the DC OPF program with respect to demand \cite[4.3]{kirschen_fundamentals_2018}. The LMPs are an essential aspect of wholesale electricity markets to ensure the generation facilities and transmission system operators can recover operating costs. Moreover, LMPs are central to recent analyses in distribution level marginal pricing \cite{papavasiliou_analysis_2018} and future distribution retail markets \cite{haider_reinventing_2021}.

There has been work to expand the assessment of the marginal change in power system factors with relation to demand. For instance, \cite{rogers_evaluation_2013} proposed an LMP based emission estimation method that contributed to the later formulation of  dynamic locational marginal emissions (LMEs) \cite{valenzuala_degleris_locational_marginal_emissions}. Assessing LMEs can help to determine the amount of emissions different generation facility distribute to individual transmission nodes. This can be helpful in deciding the trade-off between lower cost and higher emitting assets than renewable energy assets. Additionally, the LMEs can reveal inequities in the distribution of emissions associated with transmission system operation. One way to reduce nodal emissions is to reduce the nodal demand by using local renewable generation. However, this could lead to inequitable economic outcomes. For example, prosumers may defect from the power grid and force the utility to increase customer rates to recover costs from the remaining grid-connected customers. To address economic efficiency and equity, \cite{chen_retail_equity_2023} constructed a bi-level optimization problem considering cross-subsidies introduced by prosumers. Where LMEs describe the distribution of emissions, our proposed LMB metric describes the distribution of energy burden across the transmission grid. These two metrics can describe the equity of system operations and the impact of grid operators' actions on customers.

\subsubsection{Differential Optimization}

   There is a rich literature developing methods to \emph{differentiate the solutions} of convex optimization programs. A broad spectrum of applications for these methods have received considerable attention in the machine learning \cite{agrawal_differentiable_2019,pmlr-v70-amos17a}, and power systems \cite{degleris2021emissionsaware} literature, and the intersection between them \cite{Chen_POLICY_FEASIBILITY_2021}. An application particularly relevant to this work is the computation of LMEs, as in \cite{degleris2021emissionsaware}; the approach taken is similar in spirit to this work. We refer the reader to \cite{amos2023tutorial} for an extensive review of applications of these techniques in control and optimization.

        

\subsection{Contributions}
\label{sec:contributions}
Our work contributes to the intersection of the power systems and energy justice literature. To the best of our knowledge, our work provides the first representation of energy burden as an implicit function of an optimal power flow solution. This is a new contribution, that \textit{directly} connects the socioeconomic principle of energy burden to the engineering principle of optimal power flow. With this metric, we demonstrate how the change in energy burden is intrinsically integrated within the power flow, and does not need to be heuristically computed from prior power flow results. The metric reveals how changes in OPF parameters change the amount of energy burden each transmission node experiences. 

Particularly, the novelty is that the LMB is \emph{computed analytically} from the OPF solution, and is \textit{not} an additional decision variable or a post-optimization equity assessment tool. This tool can be used by system operators and regulatory bodies to determine which transmission level nodes are the most burdened by the transmission infrastructure. A more in-depth analysis can parse if the transmission lines or generating stations more heavily impact the energy burden per node. The additional detail provided by this analysis would then be able to fine-tune transmission-level infrastructure investment. This study lays the groundwork for this type of analysis, by identifying the nodes that are most burdened by the transmission infrastructure. In addition to finding the nodal location, this study provides the framework for conducting a demographic analysis of the nodes in the network. This will identify if there are any underlying equity implications within the transmission system infrastructure by not only identifying where the most burden occurs, but also who is most affected.

\subsection{Paper outline}
The rest of the paper is organized as follows. First, we develop engineering definitions for energy burden and locational marginal burden in Section \ref{sec:problem-description}. Then, avenues for how to model the retail price faced by consumers as a function of OPF dual solutions are presented in Section \ref{sec:retail-pricing}. The two concepts are then linked together in the context of a parameterized OPF problem in Section \ref{sec:computing-lmb}, where we derive closed-form analytical expressions for the LMB. The insights provided by the LMBs are then numerically demonstrated and extensively discussed in Section \ref{sec:numerical-study}, where we present simulations on the Texas A\& M synthetic Hawaii network, augmented with real census tract data. Limitations and future work are then discussed in Section \ref{sec:discussion} before we conclude in Section \ref{sec:conclusion}. Finally, Appendix \ref{apdx:dcopf-qp-details} and Appendix \ref{apdx:computing-lmb} together construct a sufficient condition for when the LMB is well-defined around an OPF solution.

\section{Problem description}
    \label{sec:problem-description}

\subsection{Energy burden: an engineering definition}
\label{sec:burden-construction}
The context of this study is an electric power system serving customers downstream of the $N$ nodes of the transmission grid. The $N$ \emph{customer aggregates} have electricity demands collected in an $N$-dimensional vector $\vd$ (in MWh), and incomes collected in an $N$-dimensional vector $\vs$ (in dollars, \$). 
Each customer pays a \emph{retail price} for their electricity, and they are collected in an $N$-dimensional vector $\vpi$. These quantities together form an \emph{energy burden function} $\vb$, which we define as follows.

\begin{definition}[Energy burden]
Given $N$ buses with incomes $\vs \in \R^n$ paying retail prices $\vpi \in \R^n$ for electricity demands $\vd \in \R^n$, the energy burden function $\vb : \R^N  \times \R^N \to  (0,1]^N$ returns a vector of the ratios between the income of each bus and the cost incurred for their electricity. The energy burden function is thus defined as:
\begin{equation}
    \label{eq:energy-burden-def}
    \vb(\vpi ; \vd) := \diag(\vd \oslash \vs) \vpi,
\end{equation}
where $\{\cdot\} \oslash \{\cdot\}$ denotes elementwise division between two vectors, and $\diag(\cdot)$ denotes a diagonal matrix with the entries of a vector argument placed along the diagonal.
\end{definition}

Importantly, we note that the burden function \eqref{eq:energy-burden-def} is a function of the retail prices $\vpi$ faced by the customers. In this work, we will treat the retail prices $\vpi$ as a function that depends on the \emph{solution to an optimization problem}; that is, the locational marginal prices (LMPs) determined by the dual optimal solution to an optimal power flow (OPF) problem. 

\subsection{Locational marginal burden}
We propose a concept that we term \emph{locational marginal burden}. The LMB describes how the energy burden of customers changes with respect to their demands. Mathematically, the LMB is defined as the entries of the vector-valued function \eqref{eq:energy-burden-def}. The proposed LMB is similar in spirit to presently existing LMP values, widely used in the U.S. to describe the price of power in the wholesale electricity market, and the LME metrics which have been proposed to assess the impact of emissions at generation facilities at individual transmission nodes.

\begin{definition}[Locational marginal burden]
    The locational marginal burden (LMB) of a bus is the change in energy burden incurred by serving one additional unit of demand at that bus.
\end{definition}

Importantly, we note that our proposed definition of LMB is inclusive of multiple levels of aggregation used to compute energy burden. This definition can describe the change in \emph{total burden} throughout the network\textemdash analogous to LMPs\textemdash in which case, the LMBs are represented as a gradient. In the more general case, it can also describe the marginal burdens between pairs of customers, i.e., the change in energy burden incurred at one bus by serving one additional MW of demand at another bus. This paper focuses on the more general case, where it is suitable to model the LMBs for a given power network as a \emph{matrix}; this matrix will be derived explicitly in Section \ref{sec:computing-lmb}.

A key challenge in constructing a notion of \emph{marginal burden} in the sense of electricity network operations is that the retail electricity price faced by consumers is a defining component of how energy burden in computed. This retail price may be a complicated function of demand, which we discuss in Section \ref{sec:retail-pricing}; moreover, these prices may be a \emph{solution to a parameterized optimization problem}.

\subsection{Combined transmission-distribution model}
\label{sec:combined-transmission-distribution-network-model}
Let $\setN = \{1,\ldots,N\}$ denote the set of buses in a transmission network, and let each transmission node $k \in \setN$ be connected to a \emph{distribution network} with a set of nodes $\setD_k = \{1,\ldots, N_k\}$. Let $\vd_k \in \R^{N_k}$ denote the demands, let $\vs_k \in \R^{N_k}$ denote the incomes, and let $\vpi_k \in \R^{N_k}$ denote the \emph{retail prices} in the distribution network $\setD_k$ downstream of transmission node $k \in \setN$. Finally, define the \emph{energy burden} corresponding to the distribution network downstream of transmission node $k$ as the  function $\vb_k : \R^{N_k} \times \R^{N_k} \to (0,1]^{N_k}$, with
\begin{equation}
\label{eq:burden-in-full-generality}
    \vb_k(\vpi_k;\vd_k) = \diag(\vd_k \oslash \vs_k) \vpi_k \quad \forall k \in \setN,
\end{equation}
where $\vd_k$ is an input parameter and the incomes $\vs_k$ are fixed. 

In our model, the retail prices $\vpi_k \in \R^{N_k}$ faced by the customers in the distribution network downstream of the transmission node $k \in \setN$ are composed of \cite{chen_retail_equity_2023}:
\begin{enumerate}
    \item The cost paid by the utility to acquire energy from the transmission system, i.e., the wholesale LMP at the transmission node $k \in \setN$, which we denote as $\vlambda_k >\vzero$, which is the dual optimal \emph{solution of an OPF problem}.
    \item The cost of operations and maintenance paid by the utility, which they wish to recover. This can include costs for network-related services, and management-related costs such as policy compliance costs, payroll, etc. which we denote as $\vomega_k >\vzero$. For simplicity, we assume this cost also incorporates network investment such as expansions, upgrades, and future investments (such as grid modernization projects or investing in advanced metering infrastructure--AMI).
    \item Optionally, the profit margin imposed, $\vrho_k : \R^{N_k} \to \R^{N_k}$, is a function of demand. We can represent this as a linear function of demand, $\vrho_k = \mPhi_k \vd_k$, where $\mPhi_k \in \R^{N_k \times N_k}$ is some profit rate matrix with positive entries.
\end{enumerate}
Then, the proposed \textit{retail prices} for the distribution network $\setD_k$ are the entries of the vector 
\begin{equation}
    \vpi_k := \vlambda_k + \vomega_k + \vrho_k,
\end{equation}
where $\vu_k := \vlambda_k + \vomega_k$ denotes the cost to the utility serving the transmission node $k$.

\section{Three retail pricing models}
\label{sec:retail-pricing}
In this section, we discuss several approximations of retail electricity prices which can be used to derive the main results of this work. While the realism of these pricing models vary, the proposed method to analyze energy burden in OPF solutions, developed in Section \ref{sec:computing-lmb}, is general to any of these models. 

Suppose that every quantity discussed above is a time varying function defined over $\setT$. Assume that it is desired to determine retail prices for each node in each distribution network using pre-collected information over a time horizon $\setT = \{ t : t_0 \leq t \leq t_f\}$. The price may be based on operational costs, profit margins, and desire for recuperating investment and upgrade costs. We further assume that each utility is not profit-seeking, such that their revenues equal their operating costs, i.e. $\vrho_k(t) = 0, \forall t \in \setT, k \in \setN$.

\subsection{Model 0: Transmission-level approximation}\label{sec:model0}
A simple model for the \emph{retail price} faced by each household in a distribution network $\setD_k$ downstream of transmission node $k \in \setN$ is simply the LMP at node $\lambda_k$,
\begin{equation}
    \vpi_k = \lambda_k \vone,
\end{equation}
where $\vone$ is an $N_k$ dimensional vector of all ones. The energy burden of each household $i \in \setD_k$ is then:
\begin{equation}
    b_{k,i} =  \frac{\lambda_k}{\mathsf{\# households}(\setD_k)}\sum_{i \in \setD_k}\frac{ d_{k,i}}{s_{k,i}}.
\end{equation}


This model makes several assumptions. It assumes that all customers in $\setD_k$ are exposed to the wholesale price of electricity, and that the utility neglects to charge for any other operational costs. This is not particularly realistic, as the retail price faced by customers is typically not the LMP itself, but a time-averaged function of the LMP at its upstream transmission node\footnote{Griddy was a power retailer in the U.S. which exposed customers directly to the wholesale electricity price. However, during the February 2021 Texas power crisis caused by winter storm Uri, customers exposed to the wholesale price saw prolonged price peaks of \$9,000 per MWh\textemdash the price cap imposed by the Electric Reliability Council of Texas (ERCOT) \cite{popik_texas_blackouts_2021}.}. Nonetheless, it is a conceptually useful model that can capture the long-run interactions between the wholesale market LMP and retail prices. Below, we outline several additional models for retail prices.

\subsection{Model 1: Representative model of present-day structures} \label{sec:model1}
Here, we propose a more representative model of today's pricing mechanism. We assume there may be multiple utilities which act as electricity service providers within their respective service regions. Each utility company has a pricing function to determine the retail costs for all its customers downstream of the transmission nodes serviced by that utility. Particularly, the LMPs and operating costs at each transmission node $k \in \setN$ are $\vlambda_k = \lambda_k \vone$, and $\vomega_k = \omega_k \vone_{N_k}$, where $\lambda_k,\omega_k \in \R$ are \emph{scalars} associated with the transmission node $k$, and $\vone_{N_k}$ is an $N_k$-dimensional vector of all ones.
\begin{assumption}
    Let $\setU_\ell \subseteq \setN$ denote the set of all transmission nodes serviced by utility $\ell = 1,\ldots,L$. Assume that the cost faced by each utility $\ell$ is fixed for all $t \in \setT$, and the utility sets time-invariant retail rates $\left\{ \vpi_k^{\ell}\right\}_{k \in \setU_\ell}$ based on information collected over the time horizon $\setT$.
\end{assumption}

Essentially, we assume that the cost to the utility and the total demand of customers is approximately the same\footnote{In general, there may be variations across seasons or other periods of the year, but the framework is general to design for multiple time horizons $\setT$.} across time intervals that are the size of $\setT$. For example, the utility may expect that their costs and the total demand of their customers may be roughly the same on a week-to-week or month-to-month basis.
The cost faced to utility $\ell=1,\ldots,L$ at  transmission node $k \in \setU_\ell$ is the \emph{scalar}
\begin{subequations}
\begin{align}
    u_k^\ell &= \mathsf{cost \ to \ purchase \ power} + \mathsf{operating \ cost}\\
    &=\int_{t \in \setT} (d_k^\ell(t)\lambda_k(t) + \omega_k^{\ell}(t)) dt,
\end{align}
\end{subequations}
where $d_k^\ell:= \sum_{i \in \setD_k} d_{k,i}^\ell$ and $\omega_k^\ell$ are the demand of customers and the operation costs of utility $\ell$ and node $k \in \setU_\ell$, respectively. We assume the retail price of the utility $\ell$ for every \emph{distribution node} in $\setD_k$ is uniform, i.e., $\vpi_k^\ell = \pi_k^\ell \vone$, where $\vone$ is a $N_k$-dimensional vector of all ones. The retail prices $\pi_k^\ell$ at transmission node $k \in \setU_\ell$ for utilities $\ell=1,\ldots,L$ are then
\begin{equation}
\label{eq:time-average}
    \pi_k^\ell = \frac{u_k^\ell}{ \int_{t \in \setT} d_k^\ell(t)} = \frac{\int_{t \in \setT} (d_k^\ell(t)\lambda_k(t) + \omega_k^{\ell}(t)) dt}{\int_{t \in \setT} d_k^\ell(t)}. 
\end{equation}

More realistically, the utility wishes to impose retail prices uniformly on all of its customers downstream of the transmission nodes within their service borders $\setU_\ell \subseteq \setN$. Thus, the uniform price will also be averaged throughout their service borders, as
\begin{equation}
\label{eq:time-spatial_average}
 \vpi_k^\ell = \pi^\ell \vone = \frac{u^\ell}{\int_{t \in \setT} d_k^\ell(t)}, \quad \forall k \in \setU_\ell,
\end{equation}
with
\begin{equation}
    u^\ell = \int_{k \in \setU_\ell} \int_{t \in \setT} \left( \lambda_k^\ell(t) d_k^\ell(t)+ \omega_k^\ell(t)\right)dt,
\end{equation}
for all $\ell=1,\ldots,L$.

Finally, for each utility $\ell=1,\ldots,L$, the energy burden function $\vb_k$ for the distribution network downstream of transmission node $k \in \setU_\ell$ is then
\begin{equation}
\vb_k =  \pi^\ell_k (\vd_k \oslash \vs_k)  \quad \forall k \in \setU_\ell, \quad \ell=1,\ldots,L.
\end{equation}
where the scalar, uniform retail price $\pi_k^\ell$ is determined using either \eqref{eq:time-average} or \eqref{eq:time-spatial_average}.

\subsection{Model 2: Futuristic distribution-level model }\label{sec:model2}
The last model is representative of a potential future retail pricing mechanism where each distribution node is assigned a distribution-level LMP (D-LMP) resulting from the solution of a distribution-level OPF. The D-LMPs are used to calculate the retail price faced by each distribution node. Note that the retail price will typically be distinct between nodes in the distribution grid, and additionally may vary over time.
In this case, the D-LMPs and operating expenses downstream of each transmission node $k \in \setN$ are vectors $\vlambda_k, \vomega_k \in \R^{N_k}$ where, for any distribution network node $i \in \setD_k$, the costs faced by the utility at some other node $j \in \setD_k \setminus \{i\}$ may be different from that of node $i$.

Therefore, the retail prices $\vpi_k$ are directly computed as a vector-valued function of time, 
\begin{equation}
\label{eq:pi-vec-dlmp-time-varying}
    \vpi_k(t) = \vlambda_k(t) + \vomega_k(t) \quad \forall k \in \setN, \quad t \in \setT.
\end{equation}
In this case, the burden faced by the customers downstream of each node $k$ is given by inputting the retail prices \eqref{eq:pi-vec-dlmp-time-varying} into the burden functions \eqref{eq:burden-in-full-generality}, and parameterizing by time $t$.

\section{Connecting energy burden and optimal power flow}\label{sec:computing-lmb}
\begin{figure}[t]
    \centering
\includegraphics[width=0.72\linewidth,keepaspectratio]{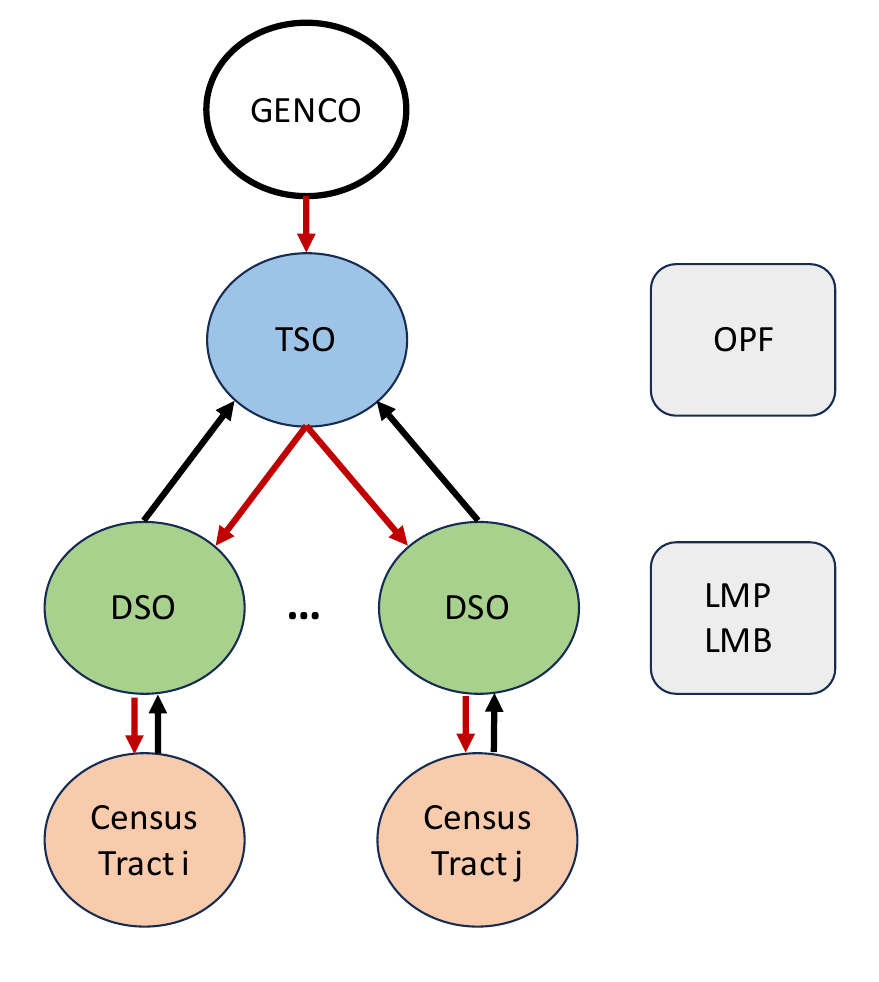}
    \caption{Illustration of the flow of information regarding power flow (red) and energy burden (black) between a generating company (GENCO), transmission system operator (TSO), distribution system operator (DSO), and represented census tracts. The outputs intended for the TSO and DSO are in the rectangles. The census tract node is an aggregate of all the census tracts downstream of the transmission node (overseen by the DSO).}
    \Description{A diagram that shows how power flow and energy burden information flows between stakeholders.}
    \label{fig:lmb-methodology-plot}
\end{figure}

We consider Model 0, where the retail prices $\vpi_k$ are functions of the \emph{solution to an optimization problem}, namely, the LMPs $\vlambda_k$. Figure \ref{fig:lmb-methodology-plot} conceptually illustrates the inherent link between OPF solutions and energy burden calculations. Therefore, the field of differentiable optimization can allow us to differentiate retail prices with respect to the fixed parameters of the problem, e.g., demand, grid constraints, and generator costs \cite{agrawal_differentiable_2019,agrawal_differentiating_2019}. We demonstrate this using the traditional DC OPF formulation to gain a theoretical understanding of how to achieve this goal; however, any convex relaxation of the power flow equations \cite{molzahn_survey_2019} can be used. Our notation is similar to \cite{valenzuala_degleris_locational_marginal_emissions}. 

\subsection{Parameterized DC OPF program}\label{sec:dcopf-parameterization}

\subsubsection{Constructing the program}
To construct the DC OPF program, we set $K$ to be the number of generators. We assume that each node has at most one generator, or equivalently, that all generators that are connected to the same node share the same cost coefficient. The program \emph{parameters} are fixed problem data collected in the vector $\vtheta := (\vd^\T,\valpha^\T,\vbeta^\T)^\T$. The parameters $\vd \in \R^{N}$ are the transmission-level active power demands $d_k = \sum_{i \in \setD_k} d_{k,i}$; $\valpha\in \R^K$ and $\vbeta \in \R^K$ are the quadratic and linear generator cost coefficients, respectively. The \emph{parameterized DC OPF program} is then
\begin{subequations}
\label{eq:static-dcopf}
    \begin{align}
    \setP(\vtheta) = \minimize_{\vg,\vp} \ &\sum_{i=1}^k \alpha_i g_i^2 + \beta_i g_i,\\
    \operatorname{s.t.}  \quad  &\mF(\mB \vg - \vd) = \vp \\
    &\vone^\top\mB \vg  = \vone^\top\vd, \label{eq:static-opf-power-balance}\\
    &-\vpubar \leq \vp \leq \vpubar \label{eq:static-opf-flow-ineq}\\
    &\vzero \leq \vg \leq \vgubar \label{eq:static-opf-gen-ineq},
    \end{align}
\end{subequations}
where $\vg \in \R^K$ are the power generation set points and $\vp \in \R^M$ are the line power flows. The matrix $\mB \in \{0,1\}^{N \times K}$ is a generator-to-node incidence matrix with entries of the form
\begin{equation}
    B_{i,j} = \begin{cases}
        1 & \mathsf{generator} \  j  \ \mathsf{at \ node} \ i\\
        0 & \mathsf{otherwise,}
    \end{cases}
\end{equation}
and the matrix $\mF \in \R^{M \times N}$ is the power transfer distribution factor (PTDF) matrix.

\subsubsection{DC OPF solution map}
We define the DC OPF solution map as follows.  Given fixed problem parameters $\vtheta$, let $\vg^*(\vtheta)$ and $\vp^*(\vtheta)$ be the primal optimal generation and branch flow variables of the DC OPF program \eqref{eq:static-dcopf}, respectively. Similarly, let $\vnu^*(\vtheta)$ and $\vmu^*(\vtheta)$ be the dual optimal variables associated with the equality and inequality constraints, respectively. 

We collect the primal and dual optimal variables in a vector that describes the DC OPF \emph{solution map}  $\vz^* : \R^{N+2K} \to \R^{4M + 3K + 1}$ as a function of the problem parameters $\vtheta$, where
\begin{equation}
    \vz^*(\vtheta) := \begin{bmatrix}
        \vg^*(\vtheta)^\T & \vp^*(\vtheta)^\T & \vmu^*(\vtheta)^\T & \vnu^*(\vtheta)^\T
    \end{bmatrix}^\T.
\end{equation}

\subsection{Deriving an LMP solution map from the OPF dual variables}
The LMPs in a transmission network are defined as the change in total operating cost incurred to serve an additional unit of demand \cite{kirschen_fundamentals_2018}. In other words, the LMPs can be obtained by taking the gradient of the Lagrangian of the DC OPF program with respect to demand \cite[4.3]{kirschen_fundamentals_2018}. See Appendix \ref{apdx:dcopf-qp-details} for an explicit derivation of the Lagrangian of the DC OPF program \eqref{eq:dcopf-lagrangian}. The LMPs can then be computed as
\begin{subequations}
\label{eq:lmps-as-gradient}
\begin{align}
    \nabla_{\vd} \c{ \mathsf{operating\ cost}} &= \nabla_{\vd} \c{l(\vz^*(\vtheta);\vd)}\\
    &= \nabla_{\vd} \left\{ -(\vnu^*(\vtheta))^\T\begin{bmatrix}
        \mF \vd\\
        \vone^\T \vd
    \end{bmatrix}\right\}\\
    &= - \begin{bmatrix}
        \mF^\T & \vone
    \end{bmatrix} \vnu^*(\vtheta).
\end{align}
\end{subequations}
As the LMPs \eqref{eq:lmps-as-gradient} depend solely on the dual variables $\vnu^*$, the LMPs at a parameterized OPF solution $\vz^*(\vtheta)$ have a \emph{solution map}   \cite{Chen2022PSCC,li_dcopf_lmp} of the form
\begin{equation}
\label{eq:lmp-solution-map}
    \vlambda(\vnu^* ; \vtheta) := - \begin{bmatrix}
        \mF^\T & \vone
    \end{bmatrix}\vnu^*(\vtheta).
\end{equation}
Considering Model 1 from Section \ref{sec:model1}, the retail prices served downstream by transmission node $i \in \setN$ can then be written as a parameterized function of $\vnu^*$ that takes the form 
\begin{equation}
        \pi_i(\vnu^*;\vtheta) = \omega_i + \lambda_i(\vnu^*(\vtheta))
        := \omega_i + \ip{\begin{bmatrix}
        \mF_i \\
        1
    \end{bmatrix}}{\vnu^*(\vtheta)},
\end{equation}
where $\mF_i$ is the $i$-th column of the PTDF matrix, and $\omega_i$ is the operating cost paid by the utility.

\subsection{Computing locational marginal burden}
\label{sec:computing-lmb-matrix}
When applying Model 0 for retail electricity pricing, it is possible to analytically derive the LMBs. As before, we consider a transmission system with $N$ buses, $M$ lines, and $K$ generators.
\begin{assumption}
    There is a single utility servicing all customers in the network, and each customer electrically connected to either transmission or distribution grid is exposed to the wholesale price of electricity (Model 0).
\end{assumption}

Under Model 0, which is described by Assumption 2, the LMB matrices describing the change in the energy burden faced at a transmission node $i$ with respect to the demand at node $j$, can be analytically computed. By applying the chain rule, product rule, and the Implicit Function Theorem (see Appendix \ref{apdx:computing-lmb} for a detailed derivation), the LMB matrix can be computed as
\begin{equation}
    \label{eq:derived-lmb}
    \boxed{\frac{\partial \vb}{\partial \vd} = -\diag(\vd \oslash \vs)\begin{bmatrix}
       \mF^\T & \vone 
    \end{bmatrix} \frac{\partial \vnu^*}{\partial \vd} + \diag\left(\vpi\left(\vnu^*\right) \oslash \vs\right)
    }
\end{equation}
where $\frac{\partial \vnu^*}{\partial \vd}$ is Jacobian of the solution of the dual optimal variables associated with the equality constraints of the DC OPF program \eqref{eq:static-dcopf} with respect to the demand. 

Note that computing $\frac{\partial \vnu^*}{\partial \vd}$ is equivalent to solving a linear system of equations, which means that the computational complexity of computing the LMB matrix \eqref{eq:derived-lmb} is at most $\setO(n^3)$. Additionally, applying results from the differentiable optimization literature allows us to establish a sufficient condition for when the LMB matrix is guaranteed to be well-defined in a region around a given OPF solution. This result is presented in Appendix \ref{apdx:computing-lmb}.



Analogously to the total operational cost used in traditional LMP calculations, let $B(\vpi;\vd) := \sum_{k \in \setN} b_k(\vpi;\vd)$ be the total energy burden throughout the network. Then, using $B$, the net marginal burden for the network can also be calculated, as an analogous quantity to traditional LMPs. The gradient of the total burden $B$ with respect to demand is $\nabla_{\vd} B = \DBDD^\T \vone$, where
\begin{equation}
\label{eq:derived-lmb-gradient}
    \nabla_{\vd} B = - \vd^\T \diag(\vs)^{-1} \begin{bmatrix}
            \mF^\T & \vone
        \end{bmatrix} \frac{\partial \vnu^*}{\partial \vd} + \vpi(\vnu^*)^\T \diag(\vs)^{-1}.
\end{equation}
We note that the net marginal burdens \eqref{eq:derived-lmb-gradient} do not capture the pairwise energy burden relationships provided by the LMB matrix \eqref{eq:derived-lmb}. This precludes analyzing the individual contribution of a node to the burden of others, which we discuss further in Section \ref{sec:burden-to-others}.


\begin{figure*}
    \centering
    \includegraphics[width=0.52\linewidth,keepaspectratio]{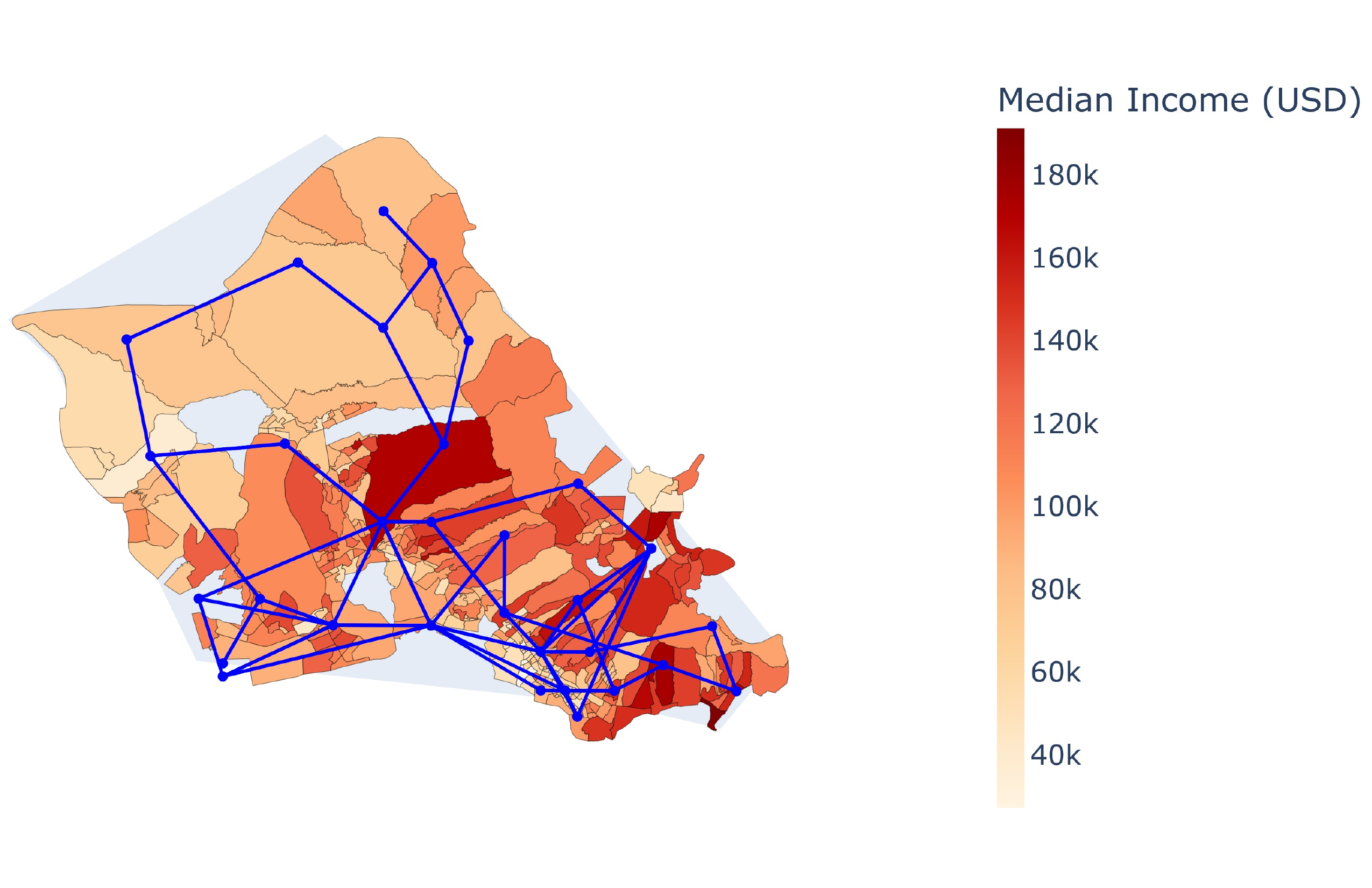}
    \caption{Mapping of publicly available household income data to the Hawaii synthetic network.}
    \Description{A map of Hawaii with income data shown on it.}
    \label{fig:hawaii-income-plot}
\end{figure*}

\section{Numerical study}
\label{sec:numerical-study}
The proposed LMB matrix \eqref{eq:derived-lmb} allows for a quantitative  assessment of how the equity of electricity rates faced by the customer(s) at each node changes\textemdash on the margin\textemdash with respect to changes in customer demand. Using real income date, we will illustrate this concept on a synthetic model of the Hawaii network. 
Figure \ref{fig:hawaii-income-plot} shows the test network overlaid with actual household income data provided by US census tract information.

\subsection{Data Sources and Processing}
\subsubsection{Network Data and Residential Demand}
We use the Texas A\&M 37-bus synthetic Hawaii network. This network geographically represents the island of Honolulu. The utility in this region is Hawaii Electric.  The data is representative of the real grid network; see \cite{birchfield_grid_2017} for additional discussion on the construction methodology used for this network. The dataset also includes estimated average residential demand across the network. 

\subsubsection{Demographic Data}
The demographic data included in this study derives the  2021 American Communities Survey (ACS) median household income estimates \cite{census2020}; it is used as the income, $s$, for each census tract in the energy burden calculation. This publicly available data is overlaid on the synthetic Hawaii network in Fig. \ref{fig:hawaii-income-plot}. 



\subsection{Energy burden computations}
We now describe how the energy burden metric and the proposed LMB metric are calculated for the synthetic Hawaii network, under Model 0.

\subsubsection{Static energy burden}

\begin{figure}
    \centering
    \includegraphics[width=\linewidth,keepaspectratio]{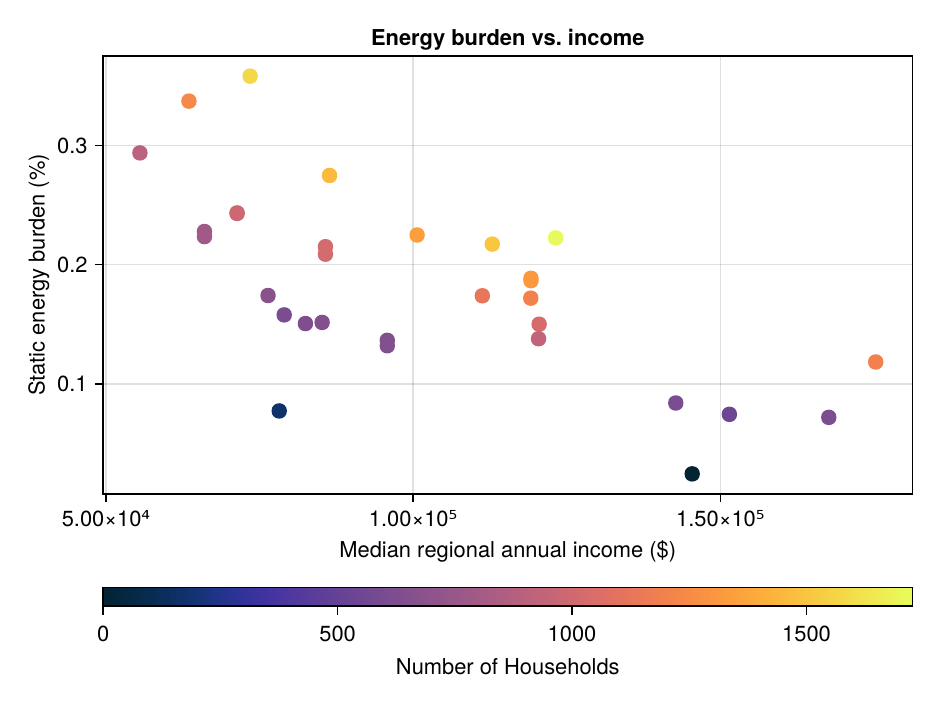}
    \caption{Static energy burden vs. mean household income in the Hawaii network, computed using \eqref{eq:burden-in-full-generality}.}
    \Description{Energy burden vs. hosehold income.}
    \label{fig:hawaii-static-burden}
\end{figure}

We first compute the static energy burden using the definition in \eqref{eq:energy-burden-def}, combined with solutions to the parameterized DC OPF program \eqref{eq:static-dcopf}. This calculation is shown in Fig. \ref{fig:hawaii-static-burden} as a function of regional income.

\subsubsection{Locational marginal burden to others}
\label{sec:burden-to-others}
Each entry $\frac{\partial b_i}{\partial d_j}$ of the matrix \eqref{eq:derived-lmb} gives us the change in energy burden faced by customer $i$ caused by the change in demand of customer $j$. Therefore, we term the sum of the off-diagonal entries for each column as the \emph{net marginal burden to others},
\begin{equation}
\label{eq:burden-to-others}
    \operatorname{\sf LMB-to-others}(i) := \sum_{j : j \neq i} \frac{\partial b_j}{\partial d_i}.
\end{equation}
In the LMB to others \eqref{eq:burden-to-others}, $\diag\left(\cdot\right)$ denotes a diagonal matrix whose entries are the diagonal elements of the matrix argument.

\begin{figure}
    \centering
    \includegraphics[width=\linewidth,keepaspectratio]{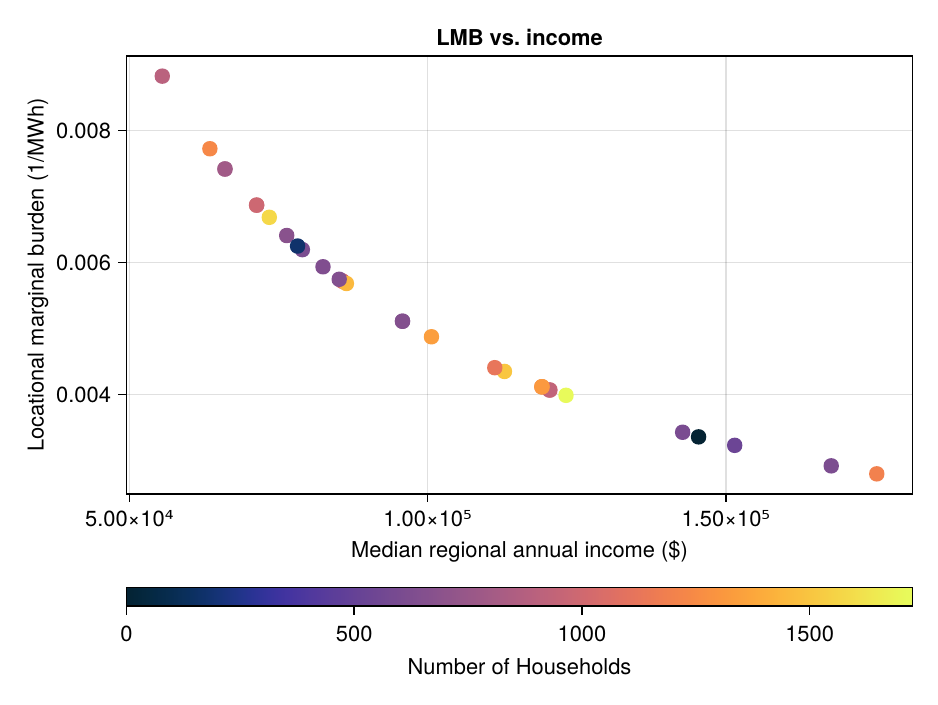}
    \caption{Diagonal entries of the LMB matrix vs. regional income in the synthetic Hawaii network with real income and census data.}
    \Description{Diagonal entries of the LMB matrix vs. regional income.}
    \label{fig:hawaii-lmb-diagonal}
\end{figure}

\begin{figure}
    \centering
    \includegraphics[width=\linewidth,keepaspectratio]{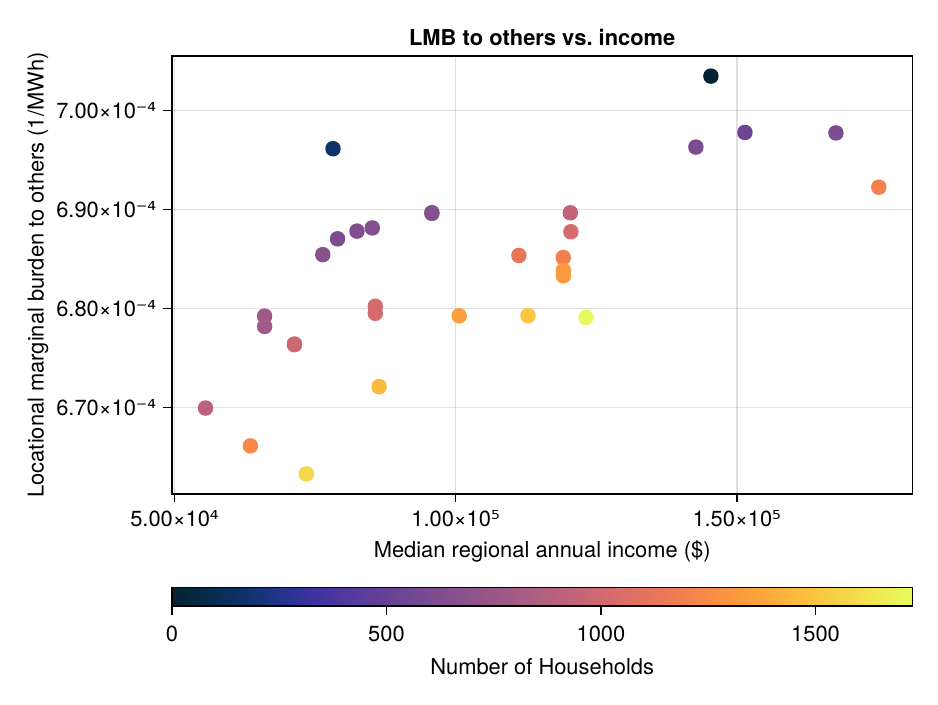}
    \caption{Net marginal burden to others, i.e., off-diagonal column sums of the LMB matrix, vs. regional income in the synthetic Hawaii network with real income and census data.}
    \Description{Net marginal burden to others vs. regional income.}
    \label{fig:hawaii-lmb-others}
\end{figure}

Plotting the diagonal entries and the off diagonal column sums of the Jacobian matrix \eqref{eq:derived-lmb} exposes the locational marginal burden, and the net marginal burden to others \eqref{eq:burden-to-others}. Fig. \ref{fig:hawaii-lmb-diagonal} and Fig. \ref{fig:hawaii-lmb-others} plot the LMB and LMB to others for the synthetic Hawaii network, respectively.

\section{Discussion}
\label{sec:discussion}

\subsection{Analyzing the synthetic Hawaii case study}
Let us first provide an example of how the LMB quantifies the equity of an OPF solution for a customer. Taking the census tract $34$ with the highest LMB of $0.09\%$, this indicates that per one additional MWh of demand at that node, there will be an increase in $0.09\%$ percent of average energy burden to all customers at that node. Note that this is an estimation of how the corresponding change in the \emph{optimal power flow solution} will impact the equity of the prices faced by customers at this node; it is not a data-driven prediction or forecast.

\subsubsection{Relationship between LMB and income}
Figure~\ref{fig:hawaii-lmb-diagonal} reveals a monotonically decreasing relationship between LMB and median annual income.  
On the other hand, there does not seem to be any correlation between the population density and the average LMB. Further analysis into the cause of the variation in nodal LMB may inform infrastructure investments regarding specific geographic regions.

\subsubsection{Relationship between the LMB to others and income}
Figure~\ref{fig:hawaii-lmb-others} plots the marginal burden to others for the synthetic Hawaii network vs. regional annual income varied by population density (color denotes the number of households). We discuss three preliminary findings based upon these results. First, we see that within a population density band (all one color) the higher income groups contribute more to the marginal burden of others. Second,  when considering regions of similar average income, areas with lower population density have higher impact on the marginal burden to others. We see this by the vertical color gradient from light to dark with increasing LMB to others, where the lighter points represent areas of high population density and the darker points represent areas of lower population density. Third, the figure shows that customers living in areas with low population density and high income, have a higher impact on LMB to others relative to customers living in high population density with low income. This is seen by inspecting the figure in the top-right corner and bottom-left corner.

These findings, in particular the second relating high and low population density within an income level, coincide with the difference between urban and rural electricity infrastructure. Urban areas are load centers with high population density, and have more transmission and distribution infrastructure, which make them more efficient and less expensive to deliver power. However, in rural areas with less transmission infrastructure, delivering power may have a higher cost. This would be reflected as higher LMP prices to transport power through transmission lines with lower capacities, resulting in line congestion and the dispatch of higher cost generation. This particular structure of our modern-day electric grid is a direct consequence of the history of electric power. Early investments into power infrastructure were entirely located in high density regions, where a single generation facility would provide power to a few hundred or few thousand customers in a minimal area, due to the costs and available technologies. The transition to utility-scale generation facilities and long-range transmission infrastructure built upon this earlier infrastructure, which connected first large cities, then smaller rural towns. Our preliminary results of the LMB on the synthetic Hawaii network may indicate the need for more grid infrastructure investments in rural communities to ensure greater pricing equity throughout the network. This introduces the discussion of evaluating and including equity in infrastructure investment funding decisions. Additionally, the first and third conclusions require further analysis with more data to make hypotheses for the cause of the results presented in this section.


\subsubsection{Intersection with Performance-based Regulatory Framework}
The proposed LMB metric can be used to inform policymakers and utility regulators about the equity of electricity pricing as it relates to the grid infrastructure. To this end, the LMB metric may have an impact on the utility business model. Traditionally, utilities operate using a cost-of-service regulatory framework, where the utility sets rates to recover the cost of supplying demand, grid investments, and other operating expenses. This may provide an incentive for utilities to invest in assets that customers may not need to increase revenue, and a disincentive to promote energy efficiency programs which reduce customer load. Recently, the framework of Performance-based Regulatory (PBR) has emerged as a new utility model which instead remunerates utilities based upon their performance in terms of reliability, resilience, and customer satisfaction, rather than by how much electricity they sell. Within the PBR framework, regulators may set performance incentive mechanisms (PIMs). Hawaii's PBR framework includes the following PIMs: supporting an accelerated renewable portfolio standard (RPS-A), interconnection approval, low-to-moderate income energy efficiency, advanced metering infrastructure (AMI) utilization, and grid services. Our LMB metric may provide an additional analysis tool to quantify how the utility achieves the equity and efficiency requirements in the low-to-moderate income energy efficiency PIM. Further, the LMB metric considers the impact of the underlying grid infrastructure on customer equity and may be helpful in considering the equity implications of grid investments and operations for the other PIMs as well.

\subsection{Limitations and future work}
One limitation in this study is the use of the demand data included in the synthetic Hawaii network. While synthetic networks are designed to be realistic, they do not represent the real system. For example, there are some census tracts modeled in the networks with zero synthetic demand; to address this, we normalized the demand by the number of households. We attempted to derive realistic demand data from a publicly available household consumption dataset. However, the solver was unable to converge with the addition of 8.28 additional MWs of demand relative to the synthetic case. This could be remedied by editing the generation capacity of the network. However, this would introduce a layer of complexity in study repeatability. Future work could seek to acquire and study actual demand data. Further, the single case study presented in this work is not indicative of broader trends. To establish clear analysis of the LMB metric, future work will consider larger integrated networks, such as the RTS-GMLC network. This network is spread across multiple states with different regulatory frameworks and different communities. 

A pertinent limitation of this work is the use of the DC OPF approximation in Section \ref{sec:dcopf-parameterization} and the LMP-centric retail pricing model described in Section \ref{sec:model0}. Although the DC approximation is widely used by transmission system operators, the inclusion of loss factors in the LMP solution map \eqref{eq:lmp-solution-map} would be a valuable extension. A significant opportunity for future work is investigating the use of more realistic retail pricing models, as discussed in Section \ref{sec:retail-pricing}. It must be noted that although we use Model 0 in the numerical study presented in this work, the framework developed in Section \ref{sec:computing-lmb} is model-agnostic; i.e., it is general to any of the proposed retail pricing models. Ongoing work by the authors will investigate the interaction of the proposed framework with the more realistic pricing models proposed in Section \ref{sec:retail-pricing}.

Applying this work to distribution networks will require AC formulations. The DC model does not consider reactive power or voltage magnitudes, which is prohibitively unrealistic in the distribution network setting; moreover, it precludes any modeling of voltage regulation and reactive power control. Note that extending the results of this paper to an AC formulation will require the use of convex relaxations of the AC power flow equations \cite{molzahn_survey_2019}, as differentiating the solution of an optimization problem requires the problem to be convex \cite{agrawal_differentiating_2019}. In the distribution network setting, using a second-order cone relaxation has been demonstrated to be effective in assessing distribution locational marginal prices \cite{papavasiliou_analysis_2018,haider_reinventing_2021}.


\section{Conclusion}
\label{sec:conclusion}
This work introduces the concept of locational marginal burden (LMB) as the sensitivity of energy burden to changes in demand in an electricity network. The key idea is to differentiate through the dual solution of an optimal power flow program, which then allows a known energy equity metric\textemdash energy burden\textemdash to be differentiated with respect to nodal demands. This concept represents a new, fundamental link between power system operations and energy equity analysis. 

Experimentally, we use synthetic Hawaii network to analyze the impact of change in demand on a network to nodal energy burden. The results suggest that the nodal LMB does not correlate with population, but with income. The correlation between the nodal LMB and income is nonlinear and inversely related to income. Additionally, when considering the impact of one node's change in demand on other nodes, the LMB to others, we observe a seemingly linear relation between increasing LMB to others and median household income. Lastly, we observe more populous census tracts distribute less burden to others when compared to less populous census tracts with similar income. 

The LMB metric can be utilized to determine the census tracts that will experience the largest change in energy burden as a result of changes in demand on the network, when utilizing the wholesale electricity rate. This can inform utility regulators or policy-makers to target infrastructure investments or capacity expansion in a manner that will reduce the marginal burden at the high LMB nodes.

\begin{acks}
This material is based upon work supported in part by the National Science Foundation Graduate Research Fellowship Program under Grant No. DGE-1650044, and NSF Grant No. 2112533. Any opinions, findings, and conclusions or recommendations expressed in this material are those of the author(s) and do not necessarily reflect the views of the National Science Foundation. 
\end{acks}

\bibliographystyle{ACM-Reference-Format}
\bibliography{refs/base,refs/extras,refs/refs}


\begin{thebibliography}{37}


\ifx \showCODEN    \undefined \def \showCODEN     #1{\unskip}     \fi
\ifx \showDOI      \undefined \def \showDOI       #1{#1}\fi
\ifx \showISBNx    \undefined \def \showISBNx     #1{\unskip}     \fi
\ifx \showISBNxiii \undefined \def \showISBNxiii  #1{\unskip}     \fi
\ifx \showISSN     \undefined \def \showISSN      #1{\unskip}     \fi
\ifx \showLCCN     \undefined \def \showLCCN      #1{\unskip}     \fi
\ifx \shownote     \undefined \def \shownote      #1{#1}          \fi
\ifx \showarticletitle \undefined \def \showarticletitle #1{#1}   \fi
\ifx \showURL      \undefined \def \showURL       {\relax}        \fi
\providecommand\bibfield[2]{#2}
\providecommand\bibinfo[2]{#2}
\providecommand\natexlab[1]{#1}
\providecommand\showeprint[2][]{arXiv:#2}

\bibitem[Agrawal et~al\mbox{.}(2019a)]%
        {agrawal_differentiable_2019}
\bibfield{author}{\bibinfo{person}{Akshay Agrawal}, \bibinfo{person}{Brandon Amos}, \bibinfo{person}{Shane Barratt}, \bibinfo{person}{Stephen Boyd}, \bibinfo{person}{Steven Diamond}, {and} \bibinfo{person}{J.~Zico Kolter}.} \bibinfo{year}{2019}\natexlab{a}.
\newblock \showarticletitle{Differentiable {Convex} {Optimization} {Layers}}. In \bibinfo{booktitle}{\emph{Advances in {Neural} {Information} {Processing} {Systems}}}, Vol.~\bibinfo{volume}{32}.
\newblock
\urldef\tempurl%
\url{https://proceedings.neurips.cc/paper_files/paper/2019/file/9ce3c52fc54362e22053399d3181c638-Paper.pdf}
\showURL{%
\tempurl}


\bibitem[Agrawal et~al\mbox{.}(2019b)]%
        {agrawal_differentiating_2019}
\bibfield{author}{\bibinfo{person}{Akshay Agrawal}, \bibinfo{person}{Shane Barratt}, \bibinfo{person}{Stephen Boyd}, \bibinfo{person}{Enzo Busseti}, {and} \bibinfo{person}{Walaa~M. Moursi}.} \bibinfo{year}{2019}\natexlab{b}.
\newblock \showarticletitle{Differentiating {Through} a {Cone} {Program}}.
\newblock \bibinfo{journal}{\emph{Journal of Applied and Numerical Optimization}} \bibinfo{volume}{1}, \bibinfo{number}{2} (\bibinfo{date}{April} \bibinfo{year}{2019}), \bibinfo{pages}{107--115}.
\newblock
\urldef\tempurl%
\url{http://arxiv.org/abs/1904.09043}
\showURL{%
\tempurl}
\newblock
\shownote{arXiv: 1904.09043}.


\bibitem[Amos(2023)]%
        {amos2023tutorial}
\bibfield{author}{\bibinfo{person}{Brandon Amos}.} \bibinfo{year}{2023}\natexlab{}.
\newblock \showarticletitle{Tutorial on Amortized Optimization}.
\newblock \bibinfo{journal}{\emph{Foundations and Trends® in Machine Learning}} \bibinfo{volume}{16}, \bibinfo{number}{5} (\bibinfo{year}{2023}), \bibinfo{pages}{592--732}.
\newblock
\showISSN{1935-8237}
\urldef\tempurl%
\url{https://doi.org/10.1561/2200000102}
\showDOI{\tempurl}


\bibitem[Amos and Kolter(2017)]%
        {pmlr-v70-amos17a}
\bibfield{author}{\bibinfo{person}{Brandon Amos} {and} \bibinfo{person}{J.~Zico Kolter}.} \bibinfo{year}{2017}\natexlab{}.
\newblock \showarticletitle{{O}pt{N}et: Differentiable Optimization as a Layer in Neural Networks}. In \bibinfo{booktitle}{\emph{Proceedings of the 34th International Conference on Machine Learning}} \emph{(\bibinfo{series}{Proceedings of Machine Learning Research}, Vol.~\bibinfo{volume}{70})}. \bibinfo{publisher}{PMLR}, \bibinfo{pages}{136--145}.
\newblock
\urldef\tempurl%
\url{https://proceedings.mlr.press/v70/amos17a.html}
\showURL{%
\tempurl}


\bibitem[Ansarin et~al\mbox{.}(2022)]%
        {ansarin_review_2022}
\bibfield{author}{\bibinfo{person}{Mohammad Ansarin}, \bibinfo{person}{Yashar Ghiassi-Farrokhfal}, \bibinfo{person}{Wolfgang Ketter}, {and} \bibinfo{person}{John Collins}.} \bibinfo{year}{2022}\natexlab{}.
\newblock \showarticletitle{A {Review} of {Equity} in {Electricity} {Tariffs} in the {Renewable} {Energy} {Era}}.
\newblock \bibinfo{journal}{\emph{Renewable and Sustainable Energy Reviews}}  \bibinfo{volume}{161} (\bibinfo{year}{2022}), \bibinfo{pages}{112333}.
\newblock
\showISSN{1364-0321}
\urldef\tempurl%
\url{https://doi.org/10.1016/j.rser.2022.112333}
\showDOI{\tempurl}


\bibitem[Balogun et~al\mbox{.}(2023)]%
        {balogun_equitable_pricing_2023}
\bibfield{author}{\bibinfo{person}{Emmanuel Balogun}, \bibinfo{person}{Sonia Martin}, \bibinfo{person}{Anthony Degleris}, {and} \bibinfo{person}{Ram Rajagopal}.} \bibinfo{year}{2023}\natexlab{}.
\newblock \showarticletitle{Equitable Dynamic Electricity Pricing via Implicitly Constrained Dual and Subgradient Methods}. In \bibinfo{booktitle}{\emph{2023 IEEE International Conference on Communications, Control, and Computing Technologies for Smart Grids (SmartGridComm)}}. \bibinfo{pages}{1--7}.
\newblock
\urldef\tempurl%
\url{https://doi.org/10.1109/SmartGridComm57358.2023.10333947}
\showDOI{\tempurl}


\bibitem[Birchfield et~al\mbox{.}(2017)]%
        {birchfield_grid_2017}
\bibfield{author}{\bibinfo{person}{Adam~B. Birchfield}, \bibinfo{person}{Ti Xu}, \bibinfo{person}{Kathleen~M. Gegner}, \bibinfo{person}{Komal~S. Shetye}, {and} \bibinfo{person}{Thomas~J. Overbye}.} \bibinfo{year}{2017}\natexlab{}.
\newblock \showarticletitle{Grid {Structural} {Characteristics} as {Validation} {Criteria} for {Synthetic} {Networks}}.
\newblock \bibinfo{journal}{\emph{IEEE Transactions on Power Systems}} \bibinfo{volume}{32}, \bibinfo{number}{4} (\bibinfo{date}{July} \bibinfo{year}{2017}), \bibinfo{pages}{3258--3265}.
\newblock
\showISSN{1558-0679}
\urldef\tempurl%
\url{https://doi.org/10.1109/TPWRS.2016.2616385}
\showDOI{\tempurl}


\bibitem[Brockway et~al\mbox{.}(2021)]%
        {brockway_inequitable_2021}
\bibfield{author}{\bibinfo{person}{Anna~M. Brockway}, \bibinfo{person}{Jennifer Conde}, {and} \bibinfo{person}{Duncan Callaway}.} \bibinfo{year}{2021}\natexlab{}.
\newblock \showarticletitle{Inequitable {Access} to {Distributed} {Energy} {Resources} {Due} to {Grid} {Infrastructure} {Limits} in {California}}.
\newblock \bibinfo{journal}{\emph{Nature Energy}} \bibinfo{volume}{6}, \bibinfo{number}{9} (\bibinfo{date}{Sept.} \bibinfo{year}{2021}), \bibinfo{pages}{892--903}.
\newblock
\showISSN{2058-7546}
\urldef\tempurl%
\url{https://doi.org/10.1038/s41560-021-00887-6}
\showDOI{\tempurl}


\bibitem[Bureau(2020)]%
        {census2020}
\bibfield{author}{\bibinfo{person}{U.S.~Census Bureau}.} \bibinfo{year}{2020}\natexlab{}.
\newblock \bibinfo{title}{Understanding and Using American Community Survey Data: What All Data Users Need to Know}.
\newblock \bibinfo{howpublished}{American Community Survey Handbooks for Data Users}.
\newblock
\urldef\tempurl%
\url{https://www.census.gov/content/dam/Census/library/publications/2020/acs/acs_geography_handbook_2020_ch01.pdf}
\showURL{%
\tempurl}


\bibitem[Charlier et~al\mbox{.}(2018)]%
        {charlier_energy_2018}
\bibfield{author}{\bibinfo{person}{Dorothée Charlier}, \bibinfo{person}{Anna Risch}, {and} \bibinfo{person}{Claire Salmon}.} \bibinfo{year}{2018}\natexlab{}.
\newblock \showarticletitle{Energy {Burden} {Alleviation} and {Greenhouse} {Gas} {Emissions} {Reduction}: {Can} {We} {Reach} {Two} {Objectives} {With} {One} {Policy}?}
\newblock \bibinfo{journal}{\emph{Ecological Economics}}  \bibinfo{volume}{143} (\bibinfo{date}{Jan.} \bibinfo{year}{2018}), \bibinfo{pages}{294--313}.
\newblock
\showISSN{0921-8009}
\urldef\tempurl%
\url{https://doi.org/10.1016/j.ecolecon.2017.07.002}
\showDOI{\tempurl}


\bibitem[Chen et~al\mbox{.}(2021)]%
        {Chen_POLICY_FEASIBILITY_2021}
\bibfield{author}{\bibinfo{person}{Bingqing Chen}, \bibinfo{person}{Priya~L. Donti}, \bibinfo{person}{Kyri Baker}, \bibinfo{person}{J.~Zico Kolter}, {and} \bibinfo{person}{Mario Berg{\'{e}}s}.} \bibinfo{year}{2021}\natexlab{}.
\newblock \showarticletitle{Enforcing Policy Feasibility Constraints through Differentiable Projection for Energy Optimization}. In \bibinfo{booktitle}{\emph{Proceedings of the Twelfth {ACM} International Conference on Future Energy Systems}}. \bibinfo{publisher}{{ACM}}.
\newblock
\urldef\tempurl%
\url{https://doi.org/10.1145/3447555.3464874}
\showDOI{\tempurl}


\bibitem[Chen et~al\mbox{.}(2022a)]%
        {chen_localized_2022}
\bibfield{author}{\bibinfo{person}{Chien-fei Chen}, \bibinfo{person}{Jimmy Feng}, \bibinfo{person}{Nikki Luke}, \bibinfo{person}{Cheng-Pin Kuo}, {and} \bibinfo{person}{Joshua~S. Fu}.} \bibinfo{year}{2022}\natexlab{a}.
\newblock \showarticletitle{Localized {Energy} {Burden}, {Concentrated} {Disadvantage}, and the {Feminization} of {Energy} {Poverty}}.
\newblock \bibinfo{journal}{\emph{iScience}} \bibinfo{volume}{25}, \bibinfo{number}{4} (\bibinfo{year}{2022}), \bibinfo{pages}{104139}.
\newblock
\showISSN{2589-0042}
\urldef\tempurl%
\url{https://doi.org/10.1016/j.isci.2022.104139}
\showDOI{\tempurl}


\bibitem[Chen et~al\mbox{.}(2023)]%
        {chen_retail_equity_2023}
\bibfield{author}{\bibinfo{person}{Yihsu Chen}, \bibinfo{person}{Andrew~L. Liu}, \bibinfo{person}{Makoto Tanaka}, {and} \bibinfo{person}{Ryuta Takashima}.} \bibinfo{year}{2023}\natexlab{}.
\newblock \showarticletitle{Optimal Retail Tariff Design With Prosumers: Pursuing Equity at the Expenses of Economic Efficiencies?}
\newblock \bibinfo{journal}{\emph{IEEE Transactions on Energy Markets, Policy and Regulation}} \bibinfo{volume}{1}, \bibinfo{number}{3} (\bibinfo{year}{2023}), \bibinfo{pages}{198--210}.
\newblock
\urldef\tempurl%
\url{https://doi.org/10.1109/TEMPR.2023.3293711}
\showDOI{\tempurl}


\bibitem[Chen et~al\mbox{.}(2022b)]%
        {Chen2022PSCC}
\bibfield{author}{\bibinfo{person}{Yize Chen}, \bibinfo{person}{Ling Zhang}, {and} \bibinfo{person}{Baosen Zhang}.} \bibinfo{year}{2022}\natexlab{b}.
\newblock \showarticletitle{Learning to Solve {DCOPF}: A Duality Approach}.
\newblock \bibinfo{journal}{\emph{Electric Power Systems Research}}  \bibinfo{volume}{213} (\bibinfo{year}{2022}), \bibinfo{pages}{108595}.
\newblock


\bibitem[Degleris et~al\mbox{.}(2021)]%
        {degleris2021emissionsaware}
\bibfield{author}{\bibinfo{person}{Anthony Degleris}, \bibinfo{person}{Lucas Fuentes}, \bibinfo{person}{Abbas El~Gamal}, {and} \bibinfo{person}{Ram Rajagopal}.} \bibinfo{year}{2021}\natexlab{}.
\newblock \showarticletitle{Emissions-Aware Electricity Network Expansion Planning via Implicit Differentiation}. In \bibinfo{booktitle}{\emph{NeurIPS 2021 Workshop on Tackling Climate Change with Machine Learning}}.
\newblock
\urldef\tempurl%
\url{https://www.climatechange.ai/papers/neurips2021/31}
\showURL{%
\tempurl}


\bibitem[for~an Energy-Efficient~Economy(2019)]%
        {ACEEEEnergyAffordability}
\bibfield{author}{\bibinfo{person}{American~Council for~an Energy-Efficient~Economy}.} \bibinfo{year}{2019}\natexlab{}.
\newblock \bibinfo{booktitle}{\emph{Understanding Energy Affordability}}.
\newblock \bibinfo{type}{Policy brief}. \bibinfo{institution}{ACEEE}, \bibinfo{address}{Washington, DC}.
\newblock
\urldef\tempurl%
\url{https://www.aceee.org/topic-brief/energy-affordability}
\showURL{%
\tempurl}


\bibitem[Haider et~al\mbox{.}(2021)]%
        {haider_reinventing_2021}
\bibfield{author}{\bibinfo{person}{Rabab Haider}, \bibinfo{person}{David D’Achiardi}, \bibinfo{person}{Venkatesh Venkataramanan}, \bibinfo{person}{Anurag Srivastava}, \bibinfo{person}{Anjan Bose}, {and} \bibinfo{person}{Anuradha~M. Annaswamy}.} \bibinfo{year}{2021}\natexlab{}.
\newblock \showarticletitle{Reinventing the {Utility} for {Distributed} {Energy} {Resources}: {A} {Proposal} for {Retail} {Electricity} {Markets}}.
\newblock \bibinfo{journal}{\emph{Advances in Applied Energy}}  \bibinfo{volume}{2} (\bibinfo{date}{May} \bibinfo{year}{2021}), \bibinfo{pages}{100026}.
\newblock
\showISSN{2666-7924}
\urldef\tempurl%
\url{https://doi.org/10.1016/j.adapen.2021.100026}
\showDOI{\tempurl}


\bibitem[Horowitz and Lave(2014)]%
        {horowitz_equity_2014}
\bibfield{author}{\bibinfo{person}{Shira Horowitz} {and} \bibinfo{person}{Lester Lave}.} \bibinfo{year}{2014}\natexlab{}.
\newblock \showarticletitle{Equity in {Residential} {Electricity} {Pricing}}.
\newblock \bibinfo{journal}{\emph{The Energy Journal}} \bibinfo{volume}{35}, \bibinfo{number}{2} (\bibinfo{date}{April} \bibinfo{year}{2014}).
\newblock
\showISSN{01956574}
\urldef\tempurl%
\url{https://doi.org/10.5547/01956574.35.2.1}
\showDOI{\tempurl}


\bibitem[Kirschen and Strbac(2018)]%
        {kirschen_fundamentals_2018}
\bibfield{author}{\bibinfo{person}{Daniel~S. Kirschen} {and} \bibinfo{person}{Goran Strbac}.} \bibinfo{year}{2018}\natexlab{}.
\newblock \bibinfo{booktitle}{\emph{Fundamentals of {Power} {System} {Economics}} (\bibinfo{edition}{2} ed.)}.
\newblock \bibinfo{publisher}{Wiley}.
\newblock
\showISBNx{978-1-119-21325-3}


\bibitem[Kody et~al\mbox{.}(2022a)]%
        {Kody2022OptimizingTI}
\bibfield{author}{\bibinfo{person}{Alyssa Kody}, \bibinfo{person}{Ryan Piansky}, {and} \bibinfo{person}{Daniel~K. Molzahn}.} \bibinfo{year}{2022}\natexlab{a}.
\newblock \showarticletitle{Optimizing Transmission Infrastructure Investments to Support Line De-energization for Mitigating Wildfire Ignition Risk}.
\newblock \bibinfo{journal}{\emph{IREP Symposium on Bulk Power System Dynamics and Control-XI}} (\bibinfo{year}{2022}).
\newblock
\urldef\tempurl%
\url{https://par.nsf.gov/biblio/10392687}
\showURL{%
\tempurl}


\bibitem[Kody et~al\mbox{.}(2022b)]%
        {Kody_sharing_2022}
\bibfield{author}{\bibinfo{person}{Alyssa Kody}, \bibinfo{person}{Amanda West}, {and} \bibinfo{person}{Daniel~K. Molzahn}.} \bibinfo{year}{2022}\natexlab{b}.
\newblock \showarticletitle{Sharing the Load: Considering Fairness in De-energization Scheduling to Mitigate Wildfire Ignition Risk using Rolling Optimization}. In \bibinfo{booktitle}{\emph{2022 IEEE 61st Conference on Decision and Control (CDC)}}. \bibinfo{pages}{5705--5712}.
\newblock
\urldef\tempurl%
\url{https://doi.org/10.1109/CDC51059.2022.9993295}
\showDOI{\tempurl}


\bibitem[Lechowicz et~al\mbox{.}(2023)]%
        {lechowicz_equitable_decarbonization_2023}
\bibfield{author}{\bibinfo{person}{Adam Lechowicz}, \bibinfo{person}{Noman Bashir}, \bibinfo{person}{John Wamburu}, \bibinfo{person}{Mohammad Hajiesmaili}, {and} \bibinfo{person}{Prashant Shenoy}.} \bibinfo{year}{2023}\natexlab{}.
\newblock \showarticletitle{Equitable Network-Aware Decarbonization of Residential Heating at City Scale}. In \bibinfo{booktitle}{\emph{Proceedings of the 14th ACM International Conference on Future Energy Systems}} (Orlando, FL, USA) \emph{(\bibinfo{series}{e-Energy '23})}. \bibinfo{publisher}{Association for Computing Machinery}, \bibinfo{address}{New York, NY, USA}, \bibinfo{pages}{1–13}.
\newblock
\showISBNx{9798400700323}
\urldef\tempurl%
\url{https://doi.org/10.1145/3575813.3576870}
\showDOI{\tempurl}


\bibitem[Li and Bo(2007)]%
        {li_dcopf_lmp}
\bibfield{author}{\bibinfo{person}{Fangxing Li} {and} \bibinfo{person}{Rui Bo}.} \bibinfo{year}{2007}\natexlab{}.
\newblock \showarticletitle{{DCOPF-Based LMP Simulation: Algorithm, Comparison With ACOPF, and Sensitivity}}.
\newblock \bibinfo{journal}{\emph{IEEE Transactions on Power Systems}} \bibinfo{volume}{22}, \bibinfo{number}{4} (\bibinfo{year}{2007}), \bibinfo{pages}{1475--1485}.
\newblock
\urldef\tempurl%
\url{https://doi.org/10.1109/TPWRS.2007.907924}
\showDOI{\tempurl}


\bibitem[Memmott et~al\mbox{.}(2021)]%
        {memmott_sociodemographic_2021}
\bibfield{author}{\bibinfo{person}{Trevor Memmott}, \bibinfo{person}{Sanya Carley}, \bibinfo{person}{Michelle Graff}, {and} \bibinfo{person}{David~M. Konisky}.} \bibinfo{year}{2021}\natexlab{}.
\newblock \showarticletitle{Sociodemographic Disparities in Energy Insecurity Among Low-Income Households Before and During the {COVID}-19 Pandemic}.
\newblock \bibinfo{journal}{\emph{Nature Energy}} \bibinfo{volume}{6}, \bibinfo{number}{2} (\bibinfo{date}{Feb.} \bibinfo{year}{2021}), \bibinfo{pages}{186--193}.
\newblock
\showISSN{2058-7546}
\urldef\tempurl%
\url{https://doi.org/10.1038/s41560-020-00763-9}
\showDOI{\tempurl}
\newblock
\shownote{Number: 2 Publisher: Nature Publishing Group}.


\bibitem[Molzahn and Hiskens(2019)]%
        {molzahn_survey_2019}
\bibfield{author}{\bibinfo{person}{Daniel~K. Molzahn} {and} \bibinfo{person}{Ian~A. Hiskens}.} \bibinfo{year}{2019}\natexlab{}.
\newblock \showarticletitle{A {Survey} of {Relaxations} and {Approximations} of the {Power} {Flow} {Equations}}.
\newblock \bibinfo{journal}{\emph{Foundations and Trends® in Electric Energy Systems}} \bibinfo{volume}{4}, \bibinfo{number}{1-2} (\bibinfo{year}{2019}), \bibinfo{pages}{1--221}.
\newblock
\showISSN{2332-6557, 2332-6565}
\urldef\tempurl%
\url{https://doi.org/10.1561/3100000012}
\showDOI{\tempurl}


\bibitem[Naughton(1986)]%
        {naughton_efficiency_1986}
\bibfield{author}{\bibinfo{person}{Michael~C. Naughton}.} \bibinfo{year}{1986}\natexlab{}.
\newblock \showarticletitle{The {Efficiency} and {Equity} {Consequences} of {Two}-{Part} {Tariffs} in {Electricity} {Pricing}}.
\newblock \bibinfo{journal}{\emph{The Review of Economics and Statistics}} \bibinfo{volume}{68}, \bibinfo{number}{3} (\bibinfo{year}{1986}), \bibinfo{pages}{406--414}.
\newblock
\showISSN{00346535, 15309142}
\urldef\tempurl%
\url{https://doi.org/10.2307/1926017}
\showDOI{\tempurl}
\newblock
\shownote{Publisher: The MIT Press}.


\bibitem[of~the President(2021)]%
        {J40_EO14008_sec223}
\bibfield{author}{\bibinfo{person}{Executive~Office of~the President}.} \bibinfo{year}{2021}\natexlab{}.
\newblock \bibinfo{title}{Tackling the {Climate} {Crisis} at {Home} and {Abroad}}.
\newblock \bibinfo{howpublished}{Executive Order No. 14008, Federal Register}.
\newblock
\urldef\tempurl%
\url{https://www.federalregister.gov/documents/2021/02/01/2021-02177/tackling-the-climate-crisis-at-home-and-abroad}
\showURL{%
\tempurl}


\bibitem[Papavasiliou(2018)]%
        {papavasiliou_analysis_2018}
\bibfield{author}{\bibinfo{person}{Anthony Papavasiliou}.} \bibinfo{year}{2018}\natexlab{}.
\newblock \showarticletitle{Analysis of {Distribution} {Locational} {Marginal} {Prices}}.
\newblock \bibinfo{journal}{\emph{IEEE Transactions on Smart Grid}} \bibinfo{volume}{9}, \bibinfo{number}{5} (\bibinfo{date}{Sept.} \bibinfo{year}{2018}), \bibinfo{pages}{4872--4882}.
\newblock
\showISSN{1949-3053, 1949-3061}
\urldef\tempurl%
\url{https://doi.org/10.1109/TSG.2017.2673860}
\showDOI{\tempurl}


\bibitem[Pittman(2023)]%
        {pittman_energy_2023}
\bibfield{author}{\bibinfo{person}{Lawrence Pittman}.} \bibinfo{year}{2023}\natexlab{}.
\newblock \showarticletitle{Energy {Justice} {Addressing} {Energy} {Burden} {Inequalities} and the {Electricity} {Grid}'s {Infrastructure} {Inequities}}.
\newblock \bibinfo{journal}{\emph{Natural Resources \& Environment}} \bibinfo{volume}{37}, \bibinfo{number}{3} (\bibinfo{year}{2023}), \bibinfo{pages}{21--25}.
\newblock
\showISSN{08823812}
\urldef\tempurl%
\url{https://www.proquest.com/docview/2778660157/abstract/E2F0BE0C359B4259PQ/1}
\showURL{%
\tempurl}


\bibitem[Roald and Andersson(2018)]%
        {roald_chance-constrained_2018}
\bibfield{author}{\bibinfo{person}{Line Roald} {and} \bibinfo{person}{Göran Andersson}.} \bibinfo{year}{2018}\natexlab{}.
\newblock \showarticletitle{Chance-{Constrained} {AC} {Optimal} {Power} {Flow}: {Reformulations} and {Efficient} {Algorithms}}.
\newblock \bibinfo{journal}{\emph{IEEE Transactions on Power Systems}} \bibinfo{volume}{33}, \bibinfo{number}{3} (\bibinfo{date}{May} \bibinfo{year}{2018}), \bibinfo{pages}{2906--2918}.
\newblock
\showISSN{1558-0679}
\urldef\tempurl%
\url{https://doi.org/10.1109/TPWRS.2017.2745410}
\showDOI{\tempurl}


\bibitem[Roald et~al\mbox{.}(2023)]%
        {roald_power_2023}
\bibfield{author}{\bibinfo{person}{Line~A. Roald}, \bibinfo{person}{David Pozo}, \bibinfo{person}{Anthony Papavasiliou}, \bibinfo{person}{Daniel~K. Molzahn}, \bibinfo{person}{Jalal Kazempour}, {and} \bibinfo{person}{Antonio Conejo}.} \bibinfo{year}{2023}\natexlab{}.
\newblock \showarticletitle{Power {Systems} {Optimization} {Under} {Uncertainty}: {A} {Review} of {Methods} and {Applications}}.
\newblock \bibinfo{journal}{\emph{Electric Power Systems Research}}  \bibinfo{volume}{214} (\bibinfo{date}{Jan.} \bibinfo{year}{2023}), \bibinfo{pages}{108725}.
\newblock
\showISSN{03787796}
\urldef\tempurl%
\url{https://doi.org/10.1016/j.epsr.2022.108725}
\showDOI{\tempurl}


\bibitem[Rogers et~al\mbox{.}(2013)]%
        {rogers_evaluation_2013}
\bibfield{author}{\bibinfo{person}{Michelle~M. Rogers}, \bibinfo{person}{Yang Wang}, \bibinfo{person}{Caisheng Wang}, \bibinfo{person}{Shawn~P. McElmurry}, {and} \bibinfo{person}{Carol~J. Miller}.} \bibinfo{year}{2013}\natexlab{}.
\newblock \showarticletitle{Evaluation of a Rapid {LMP}-Based Approach for Calculating Marginal Unit Emissions}.
\newblock \bibinfo{journal}{\emph{Applied Energy}}  \bibinfo{volume}{111} (\bibinfo{date}{Nov.} \bibinfo{year}{2013}), \bibinfo{pages}{812--820}.
\newblock
\showISSN{0306-2619}
\urldef\tempurl%
\url{https://doi.org/10.1016/j.apenergy.2013.05.057}
\showDOI{\tempurl}


\bibitem[Scheier and Kittner(2022)]%
        {scheier_measurement_2022}
\bibfield{author}{\bibinfo{person}{Eric Scheier} {and} \bibinfo{person}{Noah Kittner}.} \bibinfo{year}{2022}\natexlab{}.
\newblock \showarticletitle{A Measurement Strategy to Address Disparities Across Household Energy Burdens}.
\newblock \bibinfo{journal}{\emph{Nature Communications}} \bibinfo{volume}{13}, \bibinfo{number}{1} (\bibinfo{date}{Jan.} \bibinfo{year}{2022}), \bibinfo{pages}{288}.
\newblock
\showISSN{2041-1723}
\urldef\tempurl%
\url{https://doi.org/10.1038/s41467-021-27673-y}
\showDOI{\tempurl}


\bibitem[Sundar et~al\mbox{.}(2023)]%
        {sundar2023fairly}
\bibfield{author}{\bibinfo{person}{Kaarthik Sundar}, \bibinfo{person}{Deepjyoti Deka}, {and} \bibinfo{person}{Russell Bent}.} \bibinfo{year}{2023}\natexlab{}.
\newblock \bibinfo{title}{Fairly Extreme: Minimizing Outages Equitably}.
\newblock
\newblock
\showeprint[arxiv]{2310.01348}~[math.OC]


\bibitem[Taylor et~al\mbox{.}(2023)]%
        {taylor_managing_wildfire_equity_2023}
\bibfield{author}{\bibinfo{person}{Sofia Taylor}, \bibinfo{person}{Gabriela Setyawan}, \bibinfo{person}{Bai Cui}, \bibinfo{person}{Ahmed Zamzam}, {and} \bibinfo{person}{Line~A. Roald}.} \bibinfo{year}{2023}\natexlab{}.
\newblock \showarticletitle{Managing Wildfire Risk and Promoting Equity through Optimal Configuration of Networked Microgrids}. In \bibinfo{booktitle}{\emph{Proceedings of the 14th ACM International Conference on Future Energy Systems}} (Orlando, FL, USA) \emph{(\bibinfo{series}{e-Energy '23})}. \bibinfo{publisher}{Association for Computing Machinery}, \bibinfo{address}{New York, NY, USA}, \bibinfo{pages}{189–199}.
\newblock
\showISBNx{9798400700323}
\urldef\tempurl%
\url{https://doi.org/10.1145/3575813.3595196}
\showDOI{\tempurl}


\bibitem[Valenzuela et~al\mbox{.}(2023)]%
        {valenzuala_degleris_locational_marginal_emissions}
\bibfield{author}{\bibinfo{person}{Lucas~Fuentes Valenzuela}, \bibinfo{person}{Anthony Degleris}, \bibinfo{person}{Abbas~El Gamal}, \bibinfo{person}{Marco Pavone}, {and} \bibinfo{person}{Ram Rajagopal}.} \bibinfo{year}{2023}\natexlab{}.
\newblock \showarticletitle{Dynamic Locational Marginal Emissions via Implicit Differentiation}.
\newblock \bibinfo{journal}{\emph{IEEE Transactions on Power Systems}} (\bibinfo{year}{2023}), \bibinfo{pages}{1--11}.
\newblock
\urldef\tempurl%
\url{https://doi.org/10.1109/TPWRS.2023.3247345}
\showDOI{\tempurl}


\bibitem[Wamburu et~al\mbox{.}(2022)]%
        {wamburu_data_driven_equitable_decarbonization_2022}
\bibfield{author}{\bibinfo{person}{John Wamburu}, \bibinfo{person}{Emma Grazier}, \bibinfo{person}{David Irwin}, \bibinfo{person}{Christine Crago}, {and} \bibinfo{person}{Prashant Shenoy}.} \bibinfo{year}{2022}\natexlab{}.
\newblock \showarticletitle{Data-Driven Decarbonization of Residential Heating Systems: An Equity Perspective}. In \bibinfo{booktitle}{\emph{Proceedings of the Thirteenth ACM International Conference on Future Energy Systems}} (Virtual Event) \emph{(\bibinfo{series}{e-Energy '22})}. \bibinfo{publisher}{Association for Computing Machinery}, \bibinfo{address}{New York, NY, USA}, \bibinfo{pages}{438–439}.
\newblock
\showISBNx{9781450393973}
\urldef\tempurl%
\url{https://doi.org/10.1145/3538637.3538801}
\showDOI{\tempurl}


\end{thebibliography}


\newpage
\appendix
\section{DC OPF formulation}\label{apdx:dcopf-qp-details}
With some algebra, the parameterized DC OPF program \eqref{eq:static-dcopf} can be written as a compact general-form quadratic program (QP). To achieve this, define the constraint matrices
\begin{equation}
    \label{eq:dcopf-primal-matrices}
    \mA :=\begin{bmatrix}
            \mF \mB & - \mId_M\\
            \vone^\top \mB & \vzero_{M}^\top
        \end{bmatrix} \quad \mG :=\begin{bmatrix}
            -\mId_K & \vzero_{K\times M}\\
            \mId_K & \vzero_{K\times M}\\
            \vzero_{M\times K} & -\mId_M\\
            \vzero_{M\times K} & \mId_M
        \end{bmatrix},
\end{equation}
and the constraint vectors
\begin{equation}
    \label{eq:dcopf-primal-vecs}
    \vx := \begin{bmatrix}
        \vg\\
        \vp
    \end{bmatrix}, \ \ \vy := \begin{bmatrix}
            \mF \vd\\
            \vone^\top \vd
        \end{bmatrix}, \ \ \vh := \begin{bmatrix}
            \vzero\\
            \vgubar\\
            \vpubar\\
            \vpubar
        \end{bmatrix}.
\end{equation}
Furthermore, define the quadratic cost coefficient matrix and linear cost vector as 
\begin{equation}
\label{eq:dcopf-cost-matrix-and-vector}
    \mQ := \begin{bmatrix}
        \diag(\valpha) & \vzero_{K \times M}\\
        \vzero_{M \times K} & \tau \mId_{M \times M}
    \end{bmatrix},  \ \ \vw := \begin{bmatrix}
        \vbeta\\
        \vzero_M
    \end{bmatrix}, 
\end{equation}
where $\tau>0$ is a small regularization parameter to yield a strongly convex objective.
The program \eqref{eq:static-dcopf} can now be written as 
\begin{subequations}
    \label{eq:dcopf-standard-form}
    \begin{align}
        \setP(\vtheta) =\minimize_{\vx}\quad &\frac{1}{2}\vx^\top \mQ \vx + \vw^\top \vx,\\
        \label{eq:standard:ineq-cons}
       \mathsf{subject\ to:}\quad  &\mG \vx \preceq \vh\\
           \label{eq:standard:eq-cons}
        &\mA \vx = \vy.
    \end{align}
\end{subequations}
Let $\vmu \in \R^{2K + 2M}$ and $\vnu \in \R^{M+1}$ be the dual variables associated with the inequality constraints \eqref{eq:standard:ineq-cons} and equality constraints \eqref{eq:standard:eq-cons}, respectively. Let $\vz := (\vx^\T,\vmu^\T,\vnu^\T)^\T$ be the concatenation of the primal and dual variables of the program \eqref{eq:dcopf-standard-form}.  The Lagrangian of the program is
\begin{equation}
\label{eq:dcopf-lagrangian}
    l(\vz;\vtheta) = \frac{1}{2}\vx^\top \mQ \vx + \vw^\top \vx + \vmu^\T(\mG\vx - \vh) + \vnu^\T(\mA\vx - \vy).
\end{equation}

Next, we construct a linear system of equations that describes the Karush-Kuhn-Tucker conditions for optimality of the program \eqref{eq:dcopf-standard-form}. Let $\vz^*$ be a solution to the program \eqref{eq:dcopf-standard-form}. The stationarity, complimentary slackness, and equality constraint feasibility conditions form a system of equations $\vk(\cdot)$ whose root is $\vz^*$. It is an exercise to show that the KKT operator $\vk(\cdot) $ is then 
\begin{equation}
\label{eq:kkt-operator}
    \vk(\vz;\vtheta) =  \begin{bmatrix}
            \nabla_{\vx}\loL(\vz;\vtheta)\\
            \diag(\vmu)(\mG \vx - \vh) \\
            \mA \vx - \vy
        \end{bmatrix} = \begin{bmatrix}
             \mQ \vx + \vw + \mG^\top \vmu + \mA^\top \vnu\\
            \diag(\vmu)(\mG \vx - \vh)\\
            \mA \vx - \vy
        \end{bmatrix},
\end{equation}
which satisfies $\vk(\vz^*;\vtheta) = \vzero$.

\section{Condition for computing locational marginal burden}
\label{apdx:computing-lmb}
The key idea of this appendix is to show that the computation of the LMB matrix is dependent on the differentiability of the solution of an OPF problem, and subsequently extract the Jacobian of the dual optimal variables with respect to demand, $\frac{\partial \vnu^*}{\partial \vd}$.

\begin{proposition}[Locational marginal burden in DC OPF]
    \label{prop:existence-and-uniqueness}
    Let $\vs \in \R^n$ be the net aggregate incomes of each transmission node in an electricity network, and let $\vz^* := (\vg^*,\vp^*,\vmu^*,\vnu^*) \in \R^{3k + 4m + 1}$ be a solution to the DC OPF program \eqref{eq:static-dcopf}. Suppose that
    \begin{equation}
    \label{eq:apdx:positive-income}
        s_i > 0 \quad \forall i \in \setN,
    \end{equation}
    and the Jacobian matrix 
    \begin{equation}
    \label{eq:kkt-jacobian-matrix}
        \frac{\partial \vk}{\partial \vz}(\vz^*;\vtheta) 
        = \begin{bmatrix}
            \mQ & \mG^\T & \mA^\T\\
            \diag(\vmu^*) \mG & \diag(\mG \vx^* - \vh) & \vzero\\
            \mA & \vzero & \vzero
        \end{bmatrix},
    \end{equation}
    is non-singular, where $\mQ$, $\mA$, $\mG$, and $\vh$ are as defined in \eqref{eq:dcopf-primal-matrices}, \eqref{eq:dcopf-primal-vecs}, and \eqref{eq:dcopf-cost-matrix-and-vector}. Then the locational marginal burden matrix \eqref{eq:derived-lmb} is well-defined in a neighborhood around the OPF solution $\vz^*$.
\end{proposition}
\begin{proof}
First, observe that the LMB matrix \eqref{eq:derived-lmb} is well-defined only if condition \eqref{eq:apdx:positive-income} holds. Next, we determine the structure of the LMB matrix. Let $\ve_i$ denote the $i$-th standard basis vector, where $e_{i,i} = 1$ and $e_{i,k} = 0$ for all $k \neq i$. By applying the chain and product rule to the burden function \eqref{eq:burden-in-full-generality}, and suppressing the parametric dependence on $\vd$ for convenience, the LMBs of node $i$ with respect to demands at all nodes in the network are the entries of the following vector:
\begin{subequations}
\label{eq:lmb-matrix-rows}
    \begin{align}
        \frac{\partial b_i}{\partial \vd} &= \frac{\partial}{\partial \vd} \left\{ \frac{1}{s_i} \pi_i(\vnu^*(\vd)) d_i\right\} \\
        &= \frac{1}{s_i}\left( \pi_i(\vnu^*(\vd)) \ve_i^\T + d_i \frac{\partial }{\partial \vd} \left\{ \pi_i(\vnu^*(\vd))\right\}\right)\\
        &\overset{\eqref{eq:lmps-as-gradient}}{=} \frac{1}{s_i}\left( \pi_i(\vnu^*(\vd)) \ve_i^\T - d_i \frac{\partial }{\partial \vd} \left\{ \ip{\begin{bmatrix}
            \mF_i\\
            1
        \end{bmatrix}}{\vnu^*(\vd)}\right\}\right)\\
        &=  \frac{1}{s_i}\left( \pi_i(\vnu^*(\vd)) \ve_i^\T - d_i \left( \frac{\partial \nu_0^*}{\partial \vd} + \sum_{k=1}^m F_{k,i} \frac{\partial \nu_{k}^*}{\partial \vd}\right)\right).
    \end{align}
\end{subequations}
In the above calculation, $\mF_i \in \R^{m}$ is the $i$-th column of the PTDF matrix $\mF$, $\nu_0^*$ is the optimal power balance dual variable, and $\nu_1^*,\ldots,\nu^*_m$ are the optimal flow conservation dual variables.

Then, by stacking the rows \eqref{eq:lmb-matrix-rows}, the LMB matrix takes the form of \eqref{eq:derived-lmb}. By the Implicit Function Theorem, the Jacobian of the OPF solution $\vz^*$ with respect to the demand and cost coefficients is
\begin{equation}
\label{eq:opf-solution-jacobian}
    \frac{\partial \vz}{\partial \vtheta}(\vz^*;\vtheta) :=
    \left( \frac{\partial \vk}{\partial \vz}(\vz^*;\vtheta)\right)^{-1} \frac{\partial \vk}{\partial \vtheta}(\vz^*;\vtheta),
\end{equation}
which exists if and only if the matrix \eqref{eq:kkt-jacobian-matrix} is non-singular. The Jacobian \eqref{eq:opf-solution-jacobian} is a block matrix with the following structure:
\begin{equation}
\label{eq:opf-sol-jacobian}
\renewcommand*{\arraystretch}{1.382}
    \frac{\partial \vz}{\partial \vtheta}(\vz^*;\vtheta) = \begin{bmatrix}
    \frac{\partial \vx}{\partial \vtheta} \\ \frac{\partial \vmu}{\partial \vtheta} \\ \frac{\partial \vnu}{\partial \vtheta}
    \end{bmatrix}=\begin{bmatrix}
        \frac{\partial \vg}{\partial \valpha} & \frac{\partial \vg}{\partial \vbeta} & \frac{\partial \vg}{\partial \vd} \\
        \frac{\partial \vp}{\partial \valpha} & \frac{\partial \vp}{\partial \vbeta} & \frac{\partial \vp}{\partial \vd} \\
        \frac{\partial \vmu}{\partial \valpha} & \frac{\partial \vmu}{\partial \vbeta} & \frac{\partial \vmu}{\partial \vd} \\
        \frac{\partial \vnu}{\partial \valpha} & \frac{\partial \vnu}{\partial \vbeta} & \frac{\partial \vnu}{\partial \vd} \\
    \end{bmatrix}.
\end{equation}
Extracting the Jacobian of the dual optimal solution with respect to demand, $\frac{\partial \vnu^*}{\partial \vd}$, and plugging into \eqref{eq:derived-lmb}, yields the desired result.

\end{proof}

\end{document}